%% file: m2m_paper.tex
\documentclass[useAMS,usenatbib,onecolumn,usegraphicx]{mn2e}
\usepackage{amsmath,amssymb}
\usepackage{bm}
\usepackage{graphicx}
\usepackage{color}
\title[Reconstruction of phase space distribution functions]
{The validation of made-to-measure method for reconstruction of phase space distribution functions}
\author[H. Tagawa et al.]
{H. Tagawa$^{1,2}$\thanks{E-mail: tagawahr@nao.ac.jp }, N. Gouda$^{1,2,3}$, T. Yano$^{2,3}$, T. Hara$^{1,2}$\\
$^{1}$Department of Astronomy, Graduate school of Science, The University of Tokyo, 7-3-1 Hongo Bunkyo, Tokyo 113-0033, Japan,\\
$^{2}$National Astronomical Observatory of Japan, National Institutes of Natural Sciences, 2-21-1 Osawa, Mitaka, Tokyo
181-8588, Japan\\
$^{3}$SOKENDAI(The Graduate University for Advanced Studies), Shonan Village, Hayama, Kanagawa 240-0193 Japan}

\begin{document}

\date{17 August 2016}

\pagerange{\pageref{firstpage}--\pageref{lastpage}} \pubyear{2016}

\maketitle

\label{firstpage}

\begin{abstract}
We investigate how accurately phase space distribution functions (DFs) 
in galactic models can be reconstructed by a made-to-measure (M2M) method, 
which constructs $N$-particle models of stellar systems from photometric and various kinematic data. 
The advantage of the M2M method is that this method can be applied to various galactic models 
without assumption of the spatial symmetries of gravitational potentials adopted in galactic models, 
and furthermore, numerical calculations of the orbits of the stars 
cannot be severely constrained by the capacities of computer memories.
The M2M method has been applied to various galactic models. 
However, the degree of accuracy for the recovery of DFs 
derived by the M2M method in galactic models has never been investigated carefully. 
Therefore, we show the degree of accuracy for the recovery of the DFs 
for the anisotropic Plummer model and the axisymmetric St\"{a}ckel model, 
which have analytic solutions of the DFs. 

Furthermore, this study provides the dependence of the degree of accuracy for the recovery of the DFs 
on various parameters and a procedure adopted in this paper. 
As a result, we find that the degree of accuracy for the recovery of the DFs derived by the M2M method 
for the spherical target model is a few percent, 
and more than ten percent for the axisymmetric target model. 

\end{abstract}


\section{Introduction}\label{sec_1}

Investigating the internal dynamical structures of galaxies is important to infer their formations and evolutions. 
On the other hand, 
the phase space distribution function (DF) for the total of matters in a galaxy 
leads to physical quantities that characterize the internal dynamical structure of the galaxy 
such as the mass distribution, the gravitational potential and the velocity distribution. 
Hence it is very important to get the DFs of the total of matter that has gravity in galaxies. 
The DF of a galaxy has the following characteristics: 
galaxies can be regarded as collisionless systems 
because the thermal relaxation time of the self-gravitational system 
consisting of a large number of stars ($\gtrsim10^9$) is known to be much longer than the Hubble time. 
The DFs of such collisionless stellar systems obey the collisionless Boltzmann equation. 
In addition, galaxies are often regarded as quasi steady states in the first approximation 
since the mass and velocity distributions of a certain type of galaxies 
such as elliptical galaxies and spiral galaxies are observed to be almost independent of their ages. 
Therefore, galaxies are supposed to be quasi steady state for a long time. 
Any steady-state solution of the collisionless Boltzmann equation depends on the phase-space coordinates 
only through integrals of motion by the Jeans theorem \citep{jea15}. 
Hence the DFs of the present galaxies can be mostly represented by a few integrals of motion instead of seven variables, 
which are the phase space coordinates of positions, velocities and time. 
In this way, we can reduce the number of the variables of the DFs and so the DFs can be handled easily. 

We should get the DFs for the total of matter including the dark matter and all stars that have gravity 
to infer the actual internal dynamical structures of a galaxy. 
However, the construction of the DFs from observational data is a difficult task. 
First, this is because we cannot obtain the information of the distance and proper motion of stars 
only by the photometric and spectroscopic observations \citep{mor13}. 
These observations usually give the information of the surface density 
and the line-of-sight velocity distribution (LOSVD) as functions of position on the plane of the sky. 
However, 
for the Milky Way galaxy, 
the difficulty for the construction of the DF from this point of view can be overcome 
because future high-precision astrometric observations will give us the additional information 
such as the parallaxes and proper motions of stars. 
Furthermore, actually, the era of great Galactic astrometric survey mission in the Milky Way Galaxy is coming. 
The ongoing space satellite projects 
such as Gaia \citep[e.g.][]{per01} which was launched on December $19$, $2013$ 
and began routine operations in August, $2014$, 
and Small-JASMINE \citep[Japan Astrometry Satellite Mission for INfrared Exploration, e.g.][]{gou12} 
will bring the highly precise astrometric data. 
Combining the astrometric data with the spectroscopic observations, 
we will directly obtain the six-dimensional 
phase space coordinates of observed stars. 

Another difficulty for the construction of the DFs for the total of matter 
is caused by the fact that we can obtain the DF only for finite observed stars. 
We cannot directly get the DF for the total of matter in a galaxy 
because we cannot observe the dark matter and very faint stars. 
Therefore, it is necessary to infer the DF for the total of matter 
by the development of methods. 
We will show a method for the construction of the DFs for the total of matter 
by use of the DF for the observed stars. 
We first assume a pair of the gravitational potential 
and corresponding density distribution of the total of matter as a target galactic model. 
Second, we theoretically construct the DF for the total of matter in the assumed galactic model. 
In this paper, we call this constructed DF ``a template". 
Third we assume some other target galactic models 
that have different gravitational potentials 
and constructed the DFs in these galactic models (templates). 
The DF in the most plausible galactic model will give 
the best fit for the DF for the observed stars. 
Hence, finally, we select the best fit model of the DF (best fit template) 
by the use of statistical techniques such as the maximum likelihood method 
in the comparison of the templates with observational data. 
In this way, we can infer the DF for the total of matter 
that reflects the real galaxy with high possibility. 
Here it should be remarked that in the comparison of the template 
with the observations we should take into account the observational noise, 
selection effects of a finite sample of observed stars and so on. 
It should be stressed here that 
it is very important to construct the templates sufficiently accurately 
so that  we can infer the real DF for the total of matter with high possibility 
by the comparison with accurate observations. 
Several methods are used to construct the templates as described below.

A moment-based method solves the Jeans equation so as to find most plausible DF (template) 
which is best matched to the observed mass and velocity distribution \citep*[e.g.][]{you80,bin90,mag94,mag95,cap08,cap09}. 
To solve the Jeans equation, 
it is necessary to neglect and/or assume
higher order velocity moments. 
The main drawback of this method is 
that the positive DFs is not guaranteed. 
Furthermore this method can usually only be used for spherically symmetric models. 

A DF-based method prepares models of DFs that are functions of the integrals of motion. 
Additionally, this method constrains parameters included in the function of a DF. 
These parameters are determined so that the density and/or velocity distributions 
derived from the assumed DF can match with the observed distributions. 
This method was applied for spherical models 
\citep*{dej87,ger91a,car95}, 
axisymmetric models \citep{deh94}, 
integrable systems \citep*{dej88,hun92}, 
nearly integrable potentials \citep*{deh93,bin10}, 
and action-based models \citep{bov13,pif14,san15,tri16}. 
This method is restricted in the case that a target system has analytic integrals of motion. 
However, 
the integrals of motion cannot be obtained analytically for most systems. 
The general systems require torus construction methods 
\citep*{mcg90,mcm08,ued14} 
to construct the integrals of motion as functions of six-dimensional coordinates. 

An orbit-based method calculates a weight of each orbit 
(the occupancy ratio of a stellar orbit to all stellar orbits) 
so that the 
mass and/or velocity distributions constructed from the convolution of 
the weight of each orbit are the best fit to 
the assumed or observed mass and/or velocity distributions in a target galactic model \citep*{sch79,sch93}. 
The best fitting orbital weights represent the DF of the target galaxy. 
To calculate the weight of each orbit with good accuracy requires that 
a large number of orbits should be evolved over many orbital periods in a fixed potential of the target galaxy. 
This method is used for various galactic models \citep[e.g.][]{bos08}. 
However, the number of the orbits is severely constrained by computer memory capacity. 
If there are $N$ orbits (or particles) and $J$ observables, 
this orbit-based method has to store $O(NJ)$ variables, 
while a particle-based method shown below stores only $O(N)$ variables. 

A particle-based method (hereafter, M2M method) varies the weights of particles (stars) 
while each particle is evolved in a gravitational potential of a target galaxy, 
until the constructed mass and/or velocity distributions are 
best matched to the observed mass and velocity distribution \citep[][hereafter ST96]{sye96}. 
The advantages of this method are the absence of need of the assumption of the spatial symmetries of target galaxies, 
and the number of stellar orbits can be stored under the constraint of the capacities of computer memories 
in the M2M method is much larger than that in the orbit-based method. 
The M2M method was first applied to the Milky Way's bulge and disk in \citet{bis04}, 
and most recently applied to Milky Way in \citet{por15a} and \citet{por15b}. 
Thereafter, the M2M algorithm has been improved by 
\citet[][hereafter DL07]{del07}, \citet{deh09}, \citet{lon10}, 
\citet{hun13}. 
The M2M method is also applied to various galactic models \citep*{del08,das11,lon12,mor13}. 

In this paper, we use the M2M method so as to investigate how accurately templates 
(DFs for the total of matter in galactic models) 
can be constructed. 
Here, it is notable that the accuracies of astrometric observations will be improved before long. 
Thereby, the application of the M2M method to Gaia mock data is only tried in \citet{hun14b}. 
However, it is not clear whether the accuracies of the templates are less than 
those of the observed six-dimensional coordinates of stars. 
If the uncertainties of the templates are larger than those of the observed six-dimensional coordinates of stars, 
it is not meaningful to obtain the more accurate six-dimensional observational coordinates of stars 
to find the best fit DF of the target galaxy. 
Therefore, the accuracies of templates that are compared with observational data should be improved. 
Hence, it is important to examine the degrees of accuracy for the recovery of the DFs (templates). 
In this examination, 
as the target models, we use two analytic models that are the anisotropic Plummer model 
and the axisymmetric St\"{a}ckel model, which depends on three integrals of motion. 
The reason of choice of these models is that the DFs of these models are given analytically. 
Hence we can get the degrees of accuracy with accurate quantities 
by the comparison of the constructed template with the exact solution. 
Hitherto, the degrees of accuracy for the recovery of a DF derived by the M2M method 
were presented only for spherical target models. 
In this study, ``the solution" is also constructed by the M2M method \citep{mor12} 
and so this solution is not guaranteed to be exact. 
Thus, in this paper, we show the degrees of accuracy for the recovery of the DFs 
derived by the M2M method for two specified models 
to estimate how accurately the templates can be constructed with accurate quantities. 

Furthermore, this study provides the dependence of the degrees of accuracy 
for the recovery of the DFs on various parameters and a procedure adopted in the M2M method. 
The parameters we investigate are the number of the particles used in the M2M method (particle number), 
the number of the constraints such as the density profiles and/or velocity fields (data number), 
an initial particle distribution (initial condition), 
higher order velocity moments, 
the entropy parameter, 
and the configurations for the grids of the kinematic observable. 

This paper is organized as follows. 
In Section 2, we describe the M2M method used in this paper. 
In Section 3, we show the conditions for the construction of the DFs. 
In Section 4, we present how accurately the DFs can be reconstructed by the M2M method.  
In Section 5, we discuss the dependence of the degrees of accuracy for the recovery of the DFs. 
In Section 6, we summarize this paper.

\section{THE M2M METHOD}\label{sec_2}

The goal of the M2M method is 
evolving weights of $N$-body particles orbiting 
in a gravitational potential given as target systems or calculated self-consistently by the particle distribution \citep*{deg10,hun13} 
until the constructed mass and/or velocity distributions are best matched to 
the observed mass and velocity distribution. 
In this section, we describe the M2M algorithm used in this paper. 
More detailed descriptions for the M2M technique are written in ST96, DL07, \citet{deh09}, and \citet{lon10}. 

\subsection{THE M2M ALGORITHM}\label{subsec_the}

The observables of a target system 
characterized by the phase space DF $f(\mbox{\boldmath$z$})$ of the target system are defined by
\begin{equation}
	Y_{k,j}=\int K_{k,j}(\mbox{\boldmath$z$})f(\mbox{\boldmath$z$})\mathrm{d}^6\mbox{\boldmath$z$},
\end{equation}
where $\mbox{\boldmath$z$}=(\mbox{\boldmath$r$},\mbox{\boldmath$v$})$ are the phase space coordinates of the particles, 
and $K_{k,j}$ is known as a kernel, 
which represents the degree that an orbit at $z$ contributes to a kind of an observable $k$ at the grid $j$. 
Examples of typical observable $Y_{k,j}$ is mass distributions $M_j$, or velocity dispersions. 
The corresponding observables for the model 
that is constructed by the M2M method (model observables) are given as 
\begin{equation}
	y_{k,j}(t)=\sum_{i=1}^{N}w_{i}(t)K_{k,j}[\mbox{\boldmath$z$}_i(t)].
	\label{y_give}
\end{equation}
In the case of mass distributions, $w_{i}K_{k,j}[\mbox{\boldmath$z$}_i(t)]=M\delta_{ij}w_{i}$, 
where $M$ is the total mass of the system, 
and $\delta_{ij}$ is the selection function, 
which takes 1 if the $i$-th particle exists at the grid $j$ and takes 0 otherwise. 
Here individual particles have masses $m_i=w_i M/\sum_{i=1}^N w_i$. 
In the M2M method, the weight of $i$-th particle $w_{i}(t)$ evolves until model observables $y_{k,j}$
agree with the target observables $Y_{k,j}$. 
To reduce temporal fluctuations, 
and to increase the number of effective particles that contribute to the model observables, 
the model observables $y_{k,j}(t)$ are commonly replaced as
\begin{equation}
	\tilde y_{k,j}(t)=\alpha \int _0^\infty y_{k,j}(t-\tau)e^{-\alpha \tau}{\mathrm d}\tau,
	\label{eq_smooth}
\end{equation}
where $\alpha$ is the smoothing parameter, which controls the degree of a temporal smoothing. 
The number of effective particles is increased according to the degree of the temporal smoothing 
since the weight of each particle contributes to the backward spatial regions along its trajectory. 
This temporal smoothing makes 
the number of effective particles increases from $N$ to 
\begin{equation}
	N_\mathrm{eff}=N\frac{t_{1/2}}{\Delta t},
	\label{eff_num}
\end{equation}
where $\Delta t$ is the time step of the weight evolution equation (\ref{eq:foc}), 
and $t_{1/2}=(\mathrm{ln}2)/\alpha$ is the half lifetime of the ghost particles.

The weights of the orbital particles needs to be varied 
so as to match the modelling observables $\tilde y_{k,j}$ with the target observables $Y_{k,j}$. 
This is archived by solving the differential equation called the 'force-of-change':
\begin{equation}
\label{eq:foc}
\frac{{\mathrm d}w_i(t)}{{\mathrm d}t}=\epsilon w_i(t){\left(\mu \frac{\partial S}{\partial w_i}-\sum_k^K \sum_j^{J_k}\lambda_k \frac{K_{k,j}[\mbox{\boldmath$z$}_i(t)]}{\sigma(Y_{k,j})}\Delta_{k,j}(t)\right)},
\end{equation}
where $\epsilon$ is the parameter to control the rate of a change of the weight shown in the equation (\ref{eq:foc}), 
$\sigma(Y_{k,j})$ is the error in the target observable $Y_{k,j}$, 
$\lambda_k$ is the parameter that allows us to control the contribution of the observable $k$ to the force of change \citep{hun13}, 
$J_k$ is the number of the observable $k$, 
and $K$ is the number of the kinds of the observables. 
The entropy function $S$ is defined as
\begin{equation}
	S=-\sum_{i=1}^N w_i \mathrm{log}(w_i/\hat w_i-1)
	\label{regularization_term}
\end{equation}
\citep{mor12}, 
where $\hat{w}_i$ is called priors, and traditionally set to $\hat w_i=1/N$. 
The entropy function $S$ is used for the regularization, 
and the degree of regularization is controlled by the parameter $\mu$. 
The regularization makes the distribution of the weight smooth. 

Equation (\ref{eq:foc}) maximizes the merit function 
\begin{equation}
	F=-\frac{1}{2}\chi^2+\mu S,
	\label{eq_merit}
\end{equation}
where
\begin{equation}
\chi^2=\sum_k^K \lambda_k \chi_k^2,
	\label{eq_chi}
\end{equation}
\begin{equation}
\chi^2_k=\sum_j^{J_k}\Delta_{k,j}^2,
\end{equation}
and
\begin{equation}
\label{eq:eq1}
	\Delta_{k,j}(t)=\frac{\tilde y_{k,j}-Y_{k,j}}{\sigma(Y_{k,j})}.
\end{equation}

To avoid excessive temporal smoothing, 
ST96 indicated that the smoothing parameter $\alpha$ should satisfy $2\epsilon A<\alpha$, 
where $A$ is approximately averaged value of $\lambda_k K_{k,j}\Delta_{k,j}/\sigma(M_{j})$. 
To satisfy this relation roughly, 
$\epsilon$ is given by $\epsilon=\epsilon'\epsilon''$, 
where $\epsilon'$ is set to be $2\epsilon'<\alpha$, 
and $\epsilon''=10/\mathrm{max}_{i,j}(\lambda_k K_{k,j}\Delta_{k,j}/\sigma(Y_{k,j}))$ as DL07 and \citet{hun13}. 

\input{3condition}

\input{4result}

\input{5discussion}

\section{CONCLUSION}

We have shown the degree of accuracy for the recovery of the distribution functions (DFs) 
to investigate the validation of the M2M method. 
Hitherto, the degree of accuracy for the recovery of the DFs ($f_\mathrm{dif}$) using the M2M method 
was presented only for spherical target models. 
In this previous study, the solution is also constructed by the M2M method \citep{mor12} 
and so this solution is not guaranteed to be exact. 
In this paper, we show the degree of accuracy for the recovery of 
the mass, velocity dispersion distribution and DFs 
for the anisotropic Plummer model and the axisymmetric St\"{a}ckel model, 
which depends on three integrals of motion. 
Furthermore, we provide the dependence of $f_\mathrm{dif}$ on the several parameters. 
Consequently, our main results are summarized as follows.

\begin{itemize}
	\item 	For the isotropic spherical target model, 
		we set 
		the number of the mass constraints ($N_\mathrm{m}$) and kinematic constraints ($N_\mathrm{k}$) at $100$, 
		and the number of particles used in the M2M method ($N$) at $10^6$. 
		As a result, the average of the RMS values normalized by the target values for the mass distribution 
		($\overline{\mathrm{RMS}}(m)$) is $0.36\%$. 
		The averages of the RMS values normalized by the target values 
		for the radial 
		and tangential 
		velocity dispersion distributions are $0.811\%$ and $1.04\%$. 
		The average of the absolute values of the differences 
		between the modelling and target DF ($f_\mathrm{dif}$) is $1.55\%$. 

	\item 	For the axisymmetric St\"{a}ckel target model, 
		we set that $N_\mathrm{m}=32$, $N_\mathrm{k}=16$, and $N=10^6$. 
		As a result, $\overline{\mathrm{RMS}}(m)$ is $1.17\%$, 
		the averages of the RMS values normalized by the target values for the velocity dispersion distributions of 
		the radial, azimuthal and $z$ directions are
		$3.74\%$, $6.76\%$, and $3.29\%$, and $f_\mathrm{dif}$ is $19.9\%$. 

	\item 	We represent the dependences of $f_\mathrm{dif}$ on $N$ 
		for the spherical and the axisymmetric target models. 
		Consequently, we find that the increase of $N$ from $\sim10^6$ to $\sim10^7-10^8$
		reduces $f_\mathrm{dif}$ by a few percent. 

	\item 	We show the dependence of $f_\mathrm{dif,min}$, 
		which is $f_\mathrm{dif}$ for a sufficiently large $N$, 
		on the data number $N_\mathrm{d}$ ($N_\mathrm{m}$). 
		As a result, we give the relations as 
		$f_\mathrm{dif,min}=6.5\times10^2~N_\mathrm{d}^{-1.6}\%$ ($N_\mathrm{d}\leq80$) 
		and $f_\mathrm{dif,min}=0.80\%$ ($N_\mathrm{d}\ge80$) for the isotropic Plummer target model, 
		and $f_\mathrm{dif,min}=24.3~N_\mathrm{m}^{-0.075}\%$ 
		for the axisymmetric St\"{a}ckel model with $N_\mathrm{k}=16$. 

	\item 	The results for the isotropic spherical target model indicate that 
		$f_\mathrm{dif}$ is limited at a few percent 
		according to the particle initial condition. 
		To identify the cause of the existence of the lower limit, 
		we investigated effects as 
		the higher order velocity moments for the LOSVD, 
		the entropy parameter, 
		the temporal smoothing effect, 
		and the configuration of the kinematic observables. 
		However, the cause of the lower limits of $f_\mathrm{dif,min}$ remains uncertain. 

\end{itemize}

We have shown how accurately templates (DFs) can be reconstructed. 
Our results suggest that 
the uncertainties of the templates for the axisymmetric three integrals model ($\sim$ a few tens percent) are larger than 
those of the six-dimensional coordinates of stars that will be observed by Gaia or Small-JASMINE ($\sim$ a ten percent). 
	Note that the effects of the dust extinction may reduce the accuracy of the templates as indicated in \citet{hun14b}. 
	Furthermore, although we investigated the reconstruction for the simplistic targets models,
	the accuracy of the templates will be reduced for real galaxies, which are non-axisymmetric. 
	We will investigate the influence of such effects on the recovery of the DFs in the future. 
	Thus, since our results are thought to be problematic, 
any methods that construct the templates more accurately are desired. 


\section*{Acknowledgments}

We thank the referee for providing useful comments. 
We are also thankful to Masaki Yamaguchi for useful comments on the manuscript. 
Numerical computations and analyses were carried out on Cray XC30 and computers at Center for Computational Astrophysics, 
National Astronomical Observatory of Japan. 
This was supported by the JSPS KAKENHI Grant Number23244034(Grant-in Aid for Scientific Research (A)).

\appendix

\section[]{The distribution function of the anisotropic Plummer model}
We show the DF for the spherical anisotropic Plummer model \citep{dej86}. 
A potential-density  pair of the Plummer model can be written as
\begin{equation}
\psi(r)=1/\sqrt{1+r^2},
\end{equation}
and
\begin{equation}
\rho=\frac{3}{4\pi}\psi^5.
\end{equation}
The anisotropic model DF that corresponds to the potential-density pair is given as
\begin{equation}
\label{eq:eq87}
F_q(E,L)=\frac{3\Gamma(6-q)}{2(2\pi)^{5/2}\Gamma(q/2)}E^{7/2-q}
\mathrm{H}\left(0,q/2,9/2-q,1;\frac{L^2}{2E} \right).
\end{equation}
where $q$ is the parameter, 
\begin{eqnarray}
\mathrm{H}(a,b,c,d;x)=
 \frac{\Gamma(a+b)}{\Gamma(c-a)\Gamma(a+d)}x^a\; _{2}F_{1}(a+b,1+a-c;a+d;x) &x\leq1,\\
 \frac{\Gamma(a+b)}{\Gamma(d-b)\Gamma(b+c)}x^{-b}\; _{2}F_{1}(a+b,1+b-d;b+c;\frac{1}{x}) &x\geq1,\\
\end{eqnarray}
and
\begin{equation}
 _{2}F_{1}(a,b,c;z)=\sum_{n=0}^{\infty}\frac{(a)_n(b)_n}{(c)_n n\!}z^n.
\end{equation}
This model gives the velocity dispersions as
\begin{equation}
 \sigma_r^2=\frac{1}{6-q}\frac{1}{\sqrt{1+r^2}},
\end{equation}
and
\begin{equation}
 \sigma_{\phi}^2=\sigma_{\theta}^2=\frac{1}{6-q}
 \frac{1}{\sqrt{1+r^2}}\left(1-\frac{q}{2}\frac{r^2}{1+r^2}\right),
\end{equation}
and so the anisotropic parameter $\beta$ is represented by
\begin{equation}
\beta=1-\frac{\sigma_{\phi}^2}{\sigma_r^2}=
1-\frac{\sigma_{\theta}^2}{\sigma_r^2}=\frac{q}{2}\frac{r^2}{1+r^2}.
\end{equation}
Thus, $q$ gives an anisotropy of the model. 
If $q=0$, $q>0$, and $q<0$, 
the models are isotropic, radially anisotropic, and tangentially anisotropic \citep{dej87}.

\section{The distribution function of St\"{a}ckel models with three integrals of motion}
We show the DF for the axisymmetric St\"{a}ckel model, 
which depends on three integrals of motion \citep{dej88}. 
This model is special case that 
the DF, which depends on three integrals of motion, can be written in an analytical form.

In this model, 
the total DF is given by the sum of the DF that depends on two integrals of motion ($F_1(E,I_2)$) 
and the DF that depends on three integrals of motion ($F_2(E,I_2,I_3)$). 
$F_1(E,I_2)$ is obtained by integrating density $\rho_1$ in three dimension velocity space, 
where 
\begin{equation}
\rho_1=\rho-\rho_2
\end{equation}
$\rho$ corresponds density for the total of the model, 
and $\rho_1$ and $\rho_2$ are given by
\begin{equation}
\rho_1=\int\int\int F_1(E,I_2) \mathrm{d}^3 \mbox{\boldmath$v$},~  
\rho_2=\int\int\int F_2(E,I_2,I_3) \mathrm{d}^3 \mbox{\boldmath$v$}.
\end{equation}

We use the Kuzumin-Kutuzov model \citep{kuz62} as a potential-density pair, 
which is given by
\begin{equation}
\psi(R,z)=\frac{GM}{(R^2+z^2+a^2+c^2+2\sqrt{a^2 c^2+c^2 R^2+a^2 z^2})^{1/2}},
\end{equation}
and
\begin{equation}
\rho (R,z)=\frac{Mc^2}{4\pi} \frac{(a^2+c^2)R^2+2a^2 z^2+2a^2 c^2+a^4+3a^2\sqrt{a^2 c^2+c^2 R^2+a^2 z^2}}{(a^2 c^2+c^2 R^2+a^2 z^2)^{3/2}(R^2+z^2+a^2+c^2+2\sqrt{a^2 c^2+c^2 w^2+a^2 z^2})^{3/2}}. 
\label{kk_rho}
\end{equation}
The DF that depends on two integrals of motion and represents a potential-density pair is given as
\begin{eqnarray}
\label{eq:eq97_}
F(E,I_2)=\frac{1}{(2\pi)^{5/2}}\frac{c^2}{4a}E^{5/2}&\sum_{k=0}^\infty(k+1)\frac{\Gamma(k+5)}{k+7/2}(aE)^k\Biggl(
2 \; _3F_2\left(\frac{k}{2}+\frac{5}{2},\frac{k}{2}+3,\frac{k}{2}+\frac{1}{2};k+\frac{7}{2},\frac{1}{2};2AEL_z^2  \right)\nonumber\\
&+(k+2) \; _3F_2\left(\frac{k}{2}+\frac{5}{2},\frac{k}{2}+3,\frac{k}{2}+\frac{3}{2};k+\frac{7}{2},\frac{1}{2};4AEI_2  \right) \Biggr).
\end{eqnarray}
The model can add the DF that depends on three integrals of motion ($F_2(E,I_2,I_3)$). 
$F_2(E,I_2,I_3)$ is a room to change a velocity dispersion of the model. 
To give the total density by equation (\ref{kk_rho}),
it is required to subtract the DF that depends on two integrals of motion 
whose corresponding density distribution is same as $F_2(E,I_2,I_3)$ 
from the sum of $F_2(E,I_2,I_3)$ and $F(E,I_2)$ that is corresponding to the density of equation (\ref{kk_rho}).
. 

In these models, 
$F_2(E,I_2,I_3)$ can be given by
\begin{equation}
\label{eq:eq95_2}
F_2(E,I_2,I_3)=\sum_{l,m,n}a_{lmn}E^lI_2^m(I_2+I_3)^n,
\end{equation}
and we use the corresponding density distribution 
\begin{equation}
\rho(\lambda,\nu)=\sum_{l,m,n}a_{lmn}\rho_{lmn}(\lambda,\nu),
\end{equation}
where the third integral of motion ($I_3$) in this potential is given as 
\begin{equation}
I_3=\frac{1}{2}(L^2-2I_2)+(a^2-c^2)\left(\frac{1}{2}v_z^2-z^2\frac{G(\lambda)-G(\nu)}{\lambda-\nu}\right),
\end{equation}
and
\begin{equation}
G(\tau)=\frac{GM}{c+\sqrt{\tau}}.
\end{equation}
The density distribution that corresponds to the $F=E^lI_2^m(I_2+I_3)^n$ is given as
\begin{eqnarray}
\label{eq:eq99}
\rho_{lmn}(R,z)&=&2^{3/2-n}\sqrt{\pi}n!\Gamma(l+1)R^{2m}\psi^{l+m-n+3/2}\sum_{k=0}^{n}\frac{1}{\Gamma(3/2+m+k)}\nonumber
\\
&\times&\sum_{i=0}^{k/2}\frac{\Gamma(i+1/2)\Gamma(1/2+m+k-2i)}{i!(k-2i)!}\sum_{i_1=0}^{i}\left(\begin{array}{lcr}i\\i_1\end{array}\right)(2a)^{2i_1}\sum_{i_2=0}^{i_1}\left(\begin{array}{lcr}i_1\\i_2\end{array}\right)(-4a)^{i_2}\nonumber
\\
&\times&\sum_{i_3=0}^{k-2i}\left(\begin{array}{lcr}k-2i\\i_3\end{array}\right)(-2a)^{i_3}\psi^{2i_1-i_2+i_3}(1-AR^2\psi^2)^{(i_2+i_3)/2}\nonumber
\\
&\times&\sum_{j=0}^{n-k}\frac{\Gamma(m+n+3/2-j)}{\Gamma(l+m+n+5/2-j)}\frac{(-2c)^j}{j!(n-k-j)!}\sum_{j_1=0}^j\left(\begin{array}{lcr}j\\j_1\end{array}\right)\left(\frac{a^2-c^2}{c}\right)^{j_1}\nonumber
\\
&\times&\sum_{j_2=0}^{j_1}\left(\begin{array}{lcr}j_1\\j_2\end{array}\right)(\frac{a}{c^2}-{a^2})^{j_2}\sum_{j_3=0}^{n-j-k}\left(\begin{array}{lcr}n-j-k\\j_3\end{array}\right)[2(a^2-c^2)]^{j_3}\nonumber
\\
&\times&\sum_{j_4=0}^{j_3}\left(\begin{array}{lcr}j_3\\j_4\end{array}\right)(-a)^{j_4}\psi^{j+j_1-j_2+2j_3-j_4}(1-AR^2\psi^2)^{(j_2+j_4)/2}.
\end{eqnarray}
Moreover, equation (\ref{eq:eq99}) can be rewritten as
\begin{equation}
\rho_{lmn}(R,\psi)=\sum_{p_1=l+m-n+3/2}^{l+m+n+3/2}
\sum_{p_2=0}^{n}a_{lmn}B_{p_1,p_2}R^{2m}\psi^{p_1}(1-AR^2\psi^2)^{p_2/2}.
\end{equation}
The DF that depends on two integral of motions and corresponds to this density is given as
\begin{eqnarray}
\label{eq:eq96_}
F_{lmn}(E,I_2)=\frac{1}{\sqrt{2}\pi}\sum_{p1,p2}B_{p_1,p_2}\frac{\Gamma(p_1+1)}
{\Gamma(p_1-m-1/2)\Gamma(m+1/2)}E^{p_1-m-3/2}I_2^m\nonumber\\
 \; _3F_2\left(\frac{1+p_1}{2},1+\frac{p_1}{2},\frac{-p_2}{2};p_1-\frac{m-1}{2},\frac{1}{2}+m;4AEI_2\right).
\end{eqnarray}

$F_1(E,I_2)$ is given by equation (\ref{eq:eq97_}) minus equation (\ref{eq:eq96_}), 
and $F_2(E,L_z,I_3)$ is given by equation (\ref{eq:eq95_2}). 
Thus, a self-consistent model depending on three integrals of motion can be represented analytically.


\bsp

\label{lastpage}

\end{document}

%% file: 3condition.tex
\section{MODELLING THE ANALYTIC TARGET}\label{sec_3}

In the M2M method, each particle has its own value of the integrals of motion 
according to its initial condition (initial value of its phase space coordinate). 
Therefore, the construction of the distribution of the weights of particles $w_i$ 
corresponds to the construction of the DF. 
Since the purpose of our study is 
investigating how accurately the templates can be constructed, 
we show the degrees of accuracy for the recovery of the DFs for two analytic models. 


In this section, we describe conditions for the construction of the DFs. 
The conditions are 
target models, observables and numerical conditions. 
Also, diagnostic quantities that 
quantify the degree of accuracy for the reconstruction of the target models are described.

\subsection{Analytic target models}\label{sec_4}

The spherical anisotropic Plummer model and axisymmetric st\"{a}ckel model 
whose DFs are known analytically are used as target models. 
We describe brief characteristics of these target models. 

\subsubsection{Spherical potential}

We use the anisotropic Plummer model \citep[][also shown in appendix A]{dej86} 
as the spherical target model. 
The velocity dispersion distribution 
of this model depends on one parameter $q$ shown below. 
In addition, this model has a non-rotating pattern. 
A potential-density pair of this model is known as the Plummer model \citep{plu11}, which is given by 
\begin{equation}
\psi(r)=\frac{GM}{\sqrt{b^2+r^2}}, 
\label{plum_eq}
\end{equation}
and
\begin{equation}
	\rho(r)=\frac{3M}{4\pi b^3}(1+\frac{r^2}{b^2})^{-5/2},
\end{equation}
where $G$ is the gravitational constant, and $b$ is the scale length. We use units $M=G=b=1$.
This model provides the velocity dispersions for 
the radial direction of $\sigma_r$, the polar angle direction of $\sigma_{\theta}$, and the azimuth angle direction of $\sigma_{\phi}$ 
in the spherical coordinate as follows;
\begin{equation}
 \sigma_r^2=\frac{1}{6-q}\frac{1}{\sqrt{1+r^2}},
\end{equation}
and
\begin{equation}
 \sigma_{\phi}^2=\sigma_{\theta}^2=\frac{1}{6-q}\frac{1}{\sqrt{1+r^2}}\left(1-\frac{q}{2}\frac{r^2}{1+r^2}\right).
\end{equation}
In this paper, we set the model parameter $q$ to 
$0$ for an isotropic case, 
$0.5$ for a radially anisotropic case 
and $-0.5$ for a tangentially anisotropic case.  

\subsubsection{Axisymmetric potential}

We use the st\"{a}ckel model \citep[][also shown in the appendix B]{dej88} 
as the axisymmetric target model involving three integrals. 
This system has a non-rotating pattern, 
and has three different velocity dispersions 
for each ordinary cylindrical coordinates $R$, $\theta$, and $z$ directions at an arbitrary position. 
We assume that the system is viewed from an edge-on direction so as to investigate the cases 
that the unique recovery of the DFs is theoretically possible 
using the mass distribution and the LOSVD \citep{cap07}. 

As a potential-density pair, we use the Kuzmin-Kutuzov model \citep{kuz62}, which is given by
\begin{equation}
\psi(R,z)=\frac{GM}{(R^2+z^2+a^2+c^2+2\sqrt{a^2 c^2+c^2 R^2+a^2 z^2})^{1/2}},
\label{rho_ax_k}
\end{equation}
and
\begin{equation}
\rho (R,z)=\frac{Mc^2}{4\pi} \frac{(a^2+c^2)R^2+2a^2 z^2+2a^2 c^2+a^4+3a^2\sqrt{a^2 c^2+c^2 R^2+a^2 z^2}}
{(a^2 c^2+c^2 R^2+a^2 z^2)^{3/2}(R^2+z^2+a^2+c^2+2\sqrt{a^2 c^2+c^2 w^2+a^2 z^2})^{3/2}},
\end{equation}
where $a$ and $c$ are the model parameters. 
Here, $a+c$ determines a scale length, and $c/a$ determines the spatial configuration of this system. 
If $a>c$, the model has an oblate shape, and if $a<c$, the model has a prolate shape. 
In this study, we use the oblate model with $c/a=0.75$, and units as $M=G=a+c=1$.
The target DF is composed of the sum of two parts. 
One of the parts depends 
on two integrals of motion $F_1(E,I_2)$ and another depends on three integrals of motion $F_2(E,I_2,I_3)$, 
where the second integral of motion is given as $I_2=\frac{1}{2}L_z^2$, 
$L_z$ is the angular momentum parallel to the symmetry axis, 
the third integral of motion $I_3$ is considered as a generalization of $L^2-L_z^2$, 
and $L$ is the total angular momentum.
The part of the DF that depends on the three integrals of motion, $F_2(E,I_2,I_3)$, is given as
\begin{equation}
F_2(E,I_2,I_3)=\sum_{l,m,n}a_{lmn}E^lI_2^m(I_2+I_3)^n. 
\label{eq_f2}
\end{equation}
The term given by equation (\ref{eq_f2}) makes different 
velocity dispersions in this system. 
As the target models, we choose $F_2$ as the next two cases.
\begin{eqnarray}
	\label{ax_model}
	(a)~ F_2&=&0 \nonumber\\
	(b)~ F_2&=&-0.1E^3 (I_2+I_3)-0.05E^4 (I_2+I_3)-0.01E^4I_2 (I_2+I_3)\\
	&&+1E^5 (I_2+I_3)-4E^5I_2 (I_2+I_3)-10E^6I_2 (I_2+I_3)\nonumber 
\end{eqnarray}
In case (a), 
the radial velocity dispersions equal the $z$-axis velocity dispersions, 
and the azimuthal velocity dispersions are larger than or equal to the other velocity dispersions everywhere. 
The main difference between two cases (a) and (b) is that 
the each velocity dispersion of the case (b) 
is about $10-20$ percent larger than that of the case (a) in the central regions ($R\lesssim1.5,~z\lesssim1.5$). 
In case (b), 
the radial velocity dispersion is about $10$ percent larger than the $z$-axis velocity dispersion on in $R\gtrsim1.5$
Also, we set the model parameters to represent the non-negative density everywhere. 

\subsection{Observables}

We describe the observables that are used to construct the target models in this study. 
Normally, the phase space DF cannot be given uniquely 
only by the knowledge of the mass distribution and the potential 
except that 
it is certain that the DF depends 
only on one integral of motion for a spherical target system 
or on two integrals of motion for an axisymmetric target system. 
On the other hand, for a spherical galaxy or an axisymmetric edge-on galaxy with a given potential, 
the knowledge of the surface density and the LOSVD at every spatial position on the plane of the sky 
are sufficient for the unique recovery of the DF theoretically \citep{cap07}. 
In such cases, we can well assess how accurately the DFs (templates) are reconstructed. 
Therefore, we investigate the degrees of accuracy for the recovery of the DFs 
for the spherical target model and the axisymmetric edge-on target model 
by using the mass distribution and the LOSVD as observables.

\subsubsection{Mass distribution}

When one constructs the target mass distribution of stars by the M2M method, 
one can use the surface density or space density at some grids. 
In this paper, we use the mass distribution as observables. 
Using the mass distribution is not a special situation, 
since the mass spatial distribution can be uniquely derived from the surface density theoretically 
in the cases for the spherical and the edge-on axisymmetric target systems. 

We use the Plummer model as a spherical symmetric target model. 
As the mass observable of the spherical target model, 
we use spherical polar grids extending 
from the inner boundary $r_\mathrm{min}=0.0001$ to the outer boundary $r_\mathrm{max}=5$. 
In the case of the spherical target model, we use $b~(=1)$ in the equation (\ref{plum_eq}) 
as the units of distances such as $r_\mathrm{min}$ and $r_\mathrm{max}$. 
The outer boundary gives the maximum binding energy $E_{\mathrm{max}}=\psi(r_{\mathrm{max}})$. 
We divided radial grids into $N_\mathrm{m}$ logarithmically. 
The target mass $M_j$ on the grid $j$ is given by
\begin{equation}
	M_j=4\pi \int_{r_{j-\frac{1}{2}}}^{r_{j+\frac{1}{2}}}\rho(r) r^2 \mathrm{d}r,
	\label{eq_mass1}
\end{equation}
where 
\begin{equation}
	r_j=r_\mathrm{min}r_\mathrm{cl}^{j-\frac{1}{2}},
\end{equation}
and $r_\mathrm{cl}$ is the common logarithm, which satisfies 
\begin{equation}
	r_\mathrm{min}r_\mathrm{cl}^{N_\mathrm{m}}=r_\mathrm{max}. 
\end{equation}
Equation (\ref{eq_mass1}) is integrated by the rectangle method 
with 32 equally spaced points for each grid $j$, where $j=1,...,N_\mathrm{m}$. 
Furthermore, the density is calculated as 
\begin{equation}
	\rho(r)=2\pi \int_{-\sqrt{2(\psi-E_\mathrm{max})-v_\mathrm{T}^2}}
	^{\sqrt{2(\psi-E_\mathrm{max})-v_\mathrm{T}^2}} 
	\mathrm{d}v_{r}\int_0^{\sqrt{2(\psi-E_\mathrm{max})}} v_\mathrm{T} \mathrm{d}v_\mathrm{T} 
	f(\psi-\frac{1}{2}(v_{r}^2+v_\mathrm{T}^2),rv_\mathrm{T}),
\label{eq_mass2}
\end{equation}
where $v_{r}$ and $v_\mathrm{T}$ are the velocities 
that are parallel and perpendicular to the radial direction in the polar coordinate, respectively. 
The integral in equation (\ref{eq_mass2}) is calculated 
by the Gauss-Legendre quadrature \citep{pre92} with $16\times16$ points.

Regarding the axisymmetric target model, 
we use equally spaced grids in the meridional $(R-z)$ plane. 
The grids extend to $R_\mathrm{max}=z_\mathrm{max}=5$ with $N_\mathrm{m}\times N_\mathrm{m}$ grid points, 
and we give 16 grid points in the azimuthal direction. 
In the case of the axisymmetric target model, we use $a+c~(=1)$ in the equation (\ref{rho_ax_k}) 
as the units of distances such as $R_\mathrm{max}$ and $z_\mathrm{max}$. 
Here, we set that the DFs are truncated at $E_{\mathrm{max}}=\psi(R_{\mathrm{max}},z=0)$. 
The target mass $M_{j,l}$ on the grid $(j,l)$ is given by
\begin{equation}
	M_{j,l}=2\pi \int_{R_{j-\frac{1}{2}}}
	^{R_{j+\frac{1}{2}}} R\mathrm{d}R \int_{z_{l-\frac{1}{2}}}^{z_{l+\frac{1}{2}}} \mathrm{d}z\rho(R,z),
\label{eq_mass3}
\end{equation}
where 
\begin{equation}
	R_j=R_\mathrm{max}\times \frac{j-\frac{1}{2}}{N_\mathrm{m}},
\label{eq_mass4}
\end{equation}
and
\begin{equation}
	z_l=z_\mathrm{max}\times \frac{l-\frac{1}{2}}{N_\mathrm{m}}.
	\label{eq_mass5}
\end{equation}
The integral in equation (\ref{eq_mass3}) is integrated 
by the rectangle method with $16\times16$ equally spaced points for each grid ($j,l$), 
where $j=1,...,N_\mathrm{m}$ and $l=1,...,N_\mathrm{m}$. 
The density is calculated as 
\begin{eqnarray}
	\rho(R,z)=\int_{-\sqrt{2(\psi-E_\mathrm{max})-v_{z}^2-v_{\phi}^2}}
	^{\sqrt{2(\psi-E_\mathrm{max})-v_{z}^2-v_{\phi}^2}} &\mathrm{d}v_{R}& 
	\int_{-\sqrt{2(\psi-E_\mathrm{max})-v_{z}^2}}
	^{\sqrt{2(\psi-E_\mathrm{max})-v_z^2}} \mathrm{d}v_{\phi}\nonumber
	\\
	\times\int_{-\sqrt{2(\psi-E_\mathrm{max})}}
	^{\sqrt{2(\psi-E_\mathrm{max})}} &\mathrm{d}v_{z}& f(\psi-\frac{1}{2}(v_{R}^2+v_{\phi}^2+v_{z}^2),
	\frac{1}{2}R^2v_{\phi}^2,I_3(\mbox{\boldmath$r$},\mbox{\boldmath$v$})),
	\label{eq_mass4}
\end{eqnarray}
where $v_{R}$, $v_{\phi}$, and $v_{z}$ are 
the velocities for the radial, azimuthal, and $z$-axis directions in the cylindrical coordinate, respectively.
The integral in equation (\ref{eq_mass4}) is calculated 
by the Gauss-Legendre quadrature with $8\times8\times8$ points. 

For the uncertainties in the mass observables, 
we adopt 
$\sigma(M_j)=\sqrt{M_{j} M/N}$ for the mass grid $j$ in the case of the spherical target model, 
and $\sigma(M_{j,l})=\sqrt{M_{j,l} M/N}$ for the mass grid $j,~l$ in the case of the axisymmetric target model 
similar to DL07. 

\subsubsection{Kinematics}

We use the mass weighted Gauss-Hermite moments of the LOSVD \citep*{mar93,ger93} as kinematic target observables. 
The profile of the LOS velocity can be expressed by the Gauss-Hermite series, 
which is characterized by $V$, $\sigma$ and coefficients $h_n,~n=1,...,n_\mathrm{max}$, 
where $V$ and $\sigma$ are free parameters. 
If $V$ and $\sigma$ are equal to the parameters of the best-fitting Gaussian to the LOSVD, 
then $h_1=h_2=0$ \citep*{mar93,rix97}. 

First, we describe the processes to recover the LOSVD using the Gauss-Hermite moments. 
The mass-weighted kinematic moment is given as 
\begin{equation}
b_{n,p}\equiv m_p h_{n,p}=2\sqrt{\pi}M\sum_i\delta_{pi}u_n(\nu_{pi})w_i
\label{bnp_eq}
\end{equation}
(DL07). Here, $m_p$ is the mass in the kinematic grid $p$, 
$\delta_{pi}$ selects only particles belonging to the grid $p$, and 
\begin{equation}
\nu_{pi}=\frac{v_{y,i}-V_p}{\sigma_p},
\end{equation}
where $v_{y,i}$ is the LOS velocity of particle $i$, 
$y$ is the position in the LOS direction, 
$V_p$ and $\sigma_p$ are the best-fitting Gaussian parameters of the target LOSVD in a grid $p$, 
and the dimensionless Gauss-Hermite functions \citep{ger93} are 
\begin{equation}
u_n(\nu)=(2^{n+1}\pi n!)^{-1/2}H_n(\nu)\mathrm{exp}(-\nu^2 /2),
\end{equation}
where $H_n$ are the standard Hermite polynomials. 
\citet{mag94} indicated that 
the first order errors in $h_1$, and $h_2$ are computed from those of $V$ and $\sigma$ via 
\begin{equation}
\Delta h_1 =-\frac{1}{\sqrt{2}}\frac{\Delta V}{\sigma};\ 
\Delta h_2=-\frac{1}{\sqrt{2}}\frac{\Delta\sigma}{\sigma}.
\label{dh12}
\end{equation}
By using equation (\ref{dh12}), $V$ and $\sigma$ are iteratively varied 
until both $h_1$ and $h_2$ converge to zero \citep{rix97} so as to reduce the number of parameters. 
For the kinematic observables, the kernel of the mass-weighted higher-order moments is given as 
\begin{equation}
K_{\mbox{\boldmath$j$}i}=2\sqrt(\pi)M\delta_{pi}u_n(\nu_{pi}),\ \mbox{\boldmath$j$}=\{n,p\},
\end{equation}
and equation (\ref{eq:eq1}) is given by
\begin{equation}
	\Delta_{\mbox{\boldmath$j$}}[m_p h_{n,p}]=(b_{n,p}-B_{n,p})/\sigma(B_{n,p}), 
\end{equation}
where $B_{n,p} \equiv (m_p h_{n,p})_\mathrm{target}$ is the mass-weighted Gauss-Hermite moment of the LOSVD for the target model. 
In the modelling, the terms $m_p h_{n,p}$ are included until the $4$th order $(n=1,...,4)$.

Next, we explain configurations of kinematic observables. 
As the kinematic observables of the spherical target model, 
we use two-dimensional projected polar grids extending from the inner boundary $r_{\mathrm{proj,min}}=0.0001$ to
the outer boundary $r_{\mathrm{proj,max}}=5$. 
We divided radial grids into $N_{k}$ logarithmically. 
We calculate the LOSVD $l_v$ for the spherical target model as 
\begin{equation}
	l_v(v_{\parallel},r_\mathrm{proj})=4\pi\int_0^{y_\mathrm{max}} \mathrm{d}y 
	\int_{-\sqrt{2(\psi-E_\mathrm{max})}}^{\sqrt{2(\psi-E_\mathrm{max})}} 
	\mathrm{d}v_{\perp} f(\mbox{\boldmath$r$},\mbox{\boldmath$v$}),
\label{eq_kin1}
\end{equation}
where $v_{\parallel}$ and $v_{\perp}$ are the velocities for the parallel and perpendicular to the LOS direction, respectively, 
$r_\mathrm{proj}$ is the projected radius, 
the centroid of the target model is set to be $(y,~r_\mathrm{proj})=(0,~0)$, 
and $y_\mathrm{max}$ satisfies $y_\mathrm{max}^2+r_\mathrm{proj}^2=r_\mathrm{max}^2$. 
The integral in equation (\ref{eq_kin1}) is calculated 
by the Gauss-Legendre quadrature with $64\times64$ points. 

In the case of the axisymmetric target model, 
we use the kinematic observables on a projected $(x-z)$ grid, 
where $x$ and $z$ are the directions of the parallel to the major and the minor axis of 
a projected axisymmetric galaxy, respectively. 
Since the target model is axisymmetric and non-rotating system, 
the target model is symmetry about $x=0$ and $z=0$ plane. 
Therefore, for the reduction of the calculation time, 
we use absolute values of $x$ and $z$ coordinates when the kinematics are calculated.
The grids of kinematic observables extend out to $x_\mathrm{max}=z_\mathrm{max}=5$ 
with equally spaced $N_{k}\times N_{k}$ points. 
We calculate the LOSVD for the axisymmetric target model as 
\begin{equation}
	l_v(v_{\parallel},x,z)=2\int_0^{y_\mathrm{max}} \mathrm{d}y 
	\int_{-\sqrt{2(\psi-E_\mathrm{max})-v_{z}^2}}^{\sqrt{2(\psi-E_\mathrm{max})-v_{z}^2}} \mathrm{d}v_{x}
	\int_{-\sqrt{2(\psi-E_\mathrm{max})}}^{\sqrt{2(\psi-E_\mathrm{max})}} \mathrm{d}v_{z} 
	f(\psi-\frac{1}{2}(v_{R}^2+v_{\phi}^2+v_{z}^2),\frac{1}{2}R^2v_{\phi}^2,I_3(\mbox{\boldmath$r$},\mbox{\boldmath$v$})),
	\label{eq_kin2}
\end{equation}
where $y_\mathrm{max}$ satisfies $\psi(x,y_\mathrm{max},z)=E_\mathrm{max}$, 
$v_x$ and $v_z$ are the velocities for $x$ and $z$ directions, respectively. 
The integral in equation (\ref{eq_kin2}) is calculated 
by the Gauss-Legendre quadrature with $16\times16\times16$ points. 

In the cases of both the spherical and the axisymmetric target models, 
to obtain the LOSVD parameters $V_p$, $\sigma_p$, $h_3$, and $h_4$ by the linear fitting, 
we use the values of $l_v$ at the equally spaced 100 points of $v_{\parallel}$ for each grid $p$. 
For the uncertainties in the kinematic observables, 
we adopt $\sigma(M_{p} h_{n,{p}})=\sigma(h_n)M_c\sqrt{M_{p}/M_\mathrm{c}}$, 
where $\sigma(h_n)=0.005$ is the roughly presumed error in $h_n$ as DL07, 
$M_p$ is the mass in grid $p$, and $M_c$ is the mass in the central grid.

\subsection{Numerical condition}\label{subsec_num}

Here, we describe the setups of the reconstruction of DFs by the M2M method 
such as the initial particle distribution (initial condition), 
the scheme of an orbital integration, 
the number of the particles used in the M2M method (particle number), 
the number of the constraints for the mass and velocity distribution (data number), 
and diagnostic quantities that quantify the degrees of accuracy for the reconstruction of the target models. 

\subsubsection{Initial condition}\label{subsubsec_ini}

We explain the setting of 
the initial condition. 
As a fiducial initial condition, 
the spatial distribution is given by Hernquist mass model \citep{her90}, 
and the velocity distribution is given by the Gaussian distribution 
whose dispersion is given from solving the Jeans equations for the Hernquist potential (Hernquist with Gaussian) as \citet{lon10}. 
The Hernquist mass model is given by 
\begin{equation}
	\rho(r)=\frac{Ma_\mathrm{H}}{2\pi r (r+a_\mathrm{H})^3},
\end{equation}
where $a_\mathrm{H}$ is the scale length of this model. We set $a_\mathrm{H}$ to $1$.
The dependence of the degrees of accuracy for the recovery of the DFs on the initial conditions 
is investigated in Section \ref{sec_6}. 

\subsubsection{Orbital integration}

In the M2M method, while the weights are evolved by equation (\ref{eq:foc}), 
the positions and velocities of particles are also evolved in a given gravitational potential. 
For an orbit integration, we use the standard leap-frog scheme. 
On the other hand, the integration of force-of-change shown in equation (\ref{eq:foc}) 
is calculated using the simple Euler method. 
Because we want to investigate how accurately templates are reconstructed 
in the case that an assumed gravitational potential matches to the gravitational potential of a target galaxy, 
the gravitational potential of a target galaxy is given in the construction of the DF in this paper. 
The time steps of weights evolution according to the force-of-change shown in equation (\ref{eq:foc}) 
are set to be $2\times10^5$. 
	The $2\times10^5$ steps correspond to about 200 dynamical times at the outermost radius of $r=5$. 
We have verified that $2\times10^5$ steps 
are sufficient for the convergence of the merit function $F$ and diagnostic quantities (described below) 
to its maximum and minimum values at our parameter settings, respectively. 
Finally, the particles are evolved in the gravitational potential 
for another $10^4$ steps without evolving the weight (free evolve). 
The free evolve is proposed 
to accomplish the phase mixing for the modelling weight distribution \citep{mor12}. 

\begin{table}
	\begin{center}
	\begin{minipage}{100mm}
		\caption{
			The M2M parameter used in force of change equation (\ref{eq:foc}).
		}
		\label{table_p}
\begin{tabular}{|c|c|c|}
\hline
Parameter variable &Model Value&Variable explanation\\
\hline
$dt$& 0.01 units&Orbit integral time step\\
\hline
$\epsilon'$ &0.0125&M2M evolution rate\\
\hline
$\alpha$ & 0.02625 &Smoothing rate\\
\hline
$\mu$ & 0.0 &Entropy parameter\\
\hline
$\lambda_\mathrm{m}$ & 1.0&Mass contribution\\
\hline
$\lambda_{h_1}\sim \lambda_{h_4}$ & 0.05&Velocity contribution\\
\hline
\end{tabular}
\end{minipage}
\end{center}
\end{table}
\subsubsection{Parameter setting}

The parameters used in the force-of-change shown in equation (\ref{eq:foc}) are described in Table 1. 
The values of these parameters are similar to the values used in previous studies (e.g. DL07). 
In this paper, the regularization parameter sets to $\mu =0$ so as to consider the simple cases. 

Next, we describe the important parameters for the assessment of the validation of the M2M method. 
The first parameter is the particle number. 
We set the particle number $N$ to $10^6$, 
which is typically used by the previous studies for the M2M method. 
The second parameter is the data number. 
In the case of the spherical target model, 
we choose the data number $N_\mathrm{d}$ 
(both number of the mass grid number $N_\mathrm{m}$ and the kinematic grid number $N_\mathrm{k}$, 
$N_\mathrm{m}$=$N_\mathrm{k}$=$N_\mathrm{d}$) as 100 points. 
In the case of the axisymmetric target model, 
we choose $N_\mathrm{m}=32$, and $N_\mathrm{k}=16$. 
These parameters (the particle number and the data number) 
are fixed in Section \ref{sec_5} to elucidate the fiducial cases. 
On the other hand, we investigate the dependence of the degrees of accuracy 
for the recovery of the DFs on these parameters in Section \ref{sec_6}.

\subsubsection{Computing cost}

Here, we show the computer resources to use the M2M method with some particle numbers. 
Our M2M code is written in C and parallelized with the MPI library.
We distribute the $N$ particles evenly $N_p$ processors. 
When $N_p=100$ processors are used, 
the execution time to calculate the run with $N=10^8$ during $2\times 10^5$ step requires 
about 30 hours. 
To finish the run within 
about 30 hours, 
$N_p$ requires $\sim N/10^6$ 
since the execution time is almost proportional 
to the particle number and inversely proportional to the number of processors. 

\subsection{Diagnostic quantities}

We use two diagnostic quantities to assess 
the degrees of accuracy for the recovery of the galactic models. 
The first is the sum of the absolute value of the difference 
between the DF of the target model (target) and the particle model (modelling) weight on the cell in the integrals of motion space, 
where modelling is constructed to reproduce the target model by the M2M method. 
This quantity indicates how accurately the DFs (templates) can be reconstructed. 
Since the target DF in the integrals of motion volume need to be calculated to compare the target DF with the modelling weight, 
we divide the integrals of motion space in finite cells and link each cell to the orbit that corresponds to its centroid \citep{ven08}. 

In the case of the spherical target model, 
we use the energy ($E$) and the total angular momentum ($L$) cells $(e,l)$. 
These cells equally divide $E$ space into $n_E$ pieces, 
and $L$ space into $n_{L}$ pieces for each $E$ value. 
We use $16\times8$ $(e,l)$ cells to assess the DFs. 
The target mass weight in each integrals of motion cell for the spherical target model $W_{el}$ is represented by 
\begin{equation}
	W_{el}=\int\int_\mathrm{cell}f(E,L)\Delta V(E,L)\mathrm{d}E\mathrm{d}L ,
	\label{el_int}
\end{equation}
where $\Delta V(E,L)$ is given by 
\begin{eqnarray}
	\Delta V(E,L)&=&2\pi \int\int\int_{\Omega}\left| \frac{\partial (v_{r},v_\mathrm{T})}{\partial(E,L)}\right| v_\mathrm{T} \mathrm{d}x\mathrm{d}y\mathrm{d}z\nonumber
	\\
	&=&8\pi^2 \int_{0}^{r_\mathrm{max}}\frac{v_\mathrm{T}}{\left| rv_{r}\right|}r^2 \mathrm{d}r,
	\label{vel_int}
\end{eqnarray}
and $\Omega$ is the volume in the configuration space accessible by the bound orbit. 
In these calculations, we integrate each cell ($e,l$) in equation (\ref{el_int}) with equally spaced $8\times8$ points, 
and equation (\ref{vel_int}) with equally spaced 1024000 points by the rectangle method. 
We have verified that the interval of the integration is small enough to ignore the numerical errors. 
Using the calculated target and modelling mass weight, we give the diagnostic quantity $f_\mathrm{dif}$ as
\begin{equation}
	f_\mathrm{dif}=\frac{\sum_{e,l}|w_{el}-W_{el}|}{\sum_{e,l}W_{el}}\times100~\%,
		\label{f_sph_eq}
\end{equation}
where $w_{el}$ is the modelling mass weight that is calculated by the M2M method on the $(e,l)$ cells.

In the cases of the axisymmetric target model, 
we use $E$, the angular momentum ($L_z$) and the third integral of motion ($I_3$) cells $(e,l_z,i_3)$. 
These cells equally divide $E$ space into $n_E$ pieces, 
$L_z$ space into $n_{L_z}$ pieces for each $E$ value, 
and $I_3$ space into $n_{I_3}$ pieces for each $E$ and $L_z$ value. 
We use $16\times8\times8$ $(e,l_z,i_3)$ cells to assess the axisymmetric DFs. 
The target mass weight in each integrals of motion cell for the axisymmetric target model $W_{el_zi_3}(e=1,...,n_E,l_z=1,...,n_{L_z},i_3=1,...,n_{I_3})$ is represented by 
\begin{equation}
	W_{el_z i_3}=\int\int\int_\mathrm{cell}f(E,L_z,I_3)\Delta V(E,L_z,I_3)\mathrm{d}E\mathrm{d}L_z\mathrm{d}I_3 ,
	\label{eli_int}
\end{equation}
where $\Delta V(E,L_z,I_3)$ is given by 
\begin{eqnarray}
	\Delta V(E,L_z,I_3)&=&\int\int\int_{\Omega}	
	\left| \frac{\partial (v_x,v_y,v_z)}{\partial(E,L_z,I_3)}\right| \mathrm{d}x\mathrm{d}y\mathrm{d}z\nonumber
	\\
	&=&\frac{4\pi}{\left| L_z \right|}\int_{c^2}
	^{\nu_\mathrm{max}}\int_{\lambda_\mathrm{min}}^{\lambda_\mathrm{max}}
	\frac{(\nu-\lambda)}{(\lambda-a^2)(\lambda-c^2)(\nu-a^2)(\nu-c^2)}
	\sqrt{\frac{(\lambda-a^2)(\nu-a^2)}
	{[E-V_\mathrm{eff}(\lambda)][E-V_\mathrm{eff}(\nu)]}}\mathrm{d}\lambda \mathrm{d}\nu.
	\label{veli_int}
\end{eqnarray}
In these calculations, we integrate each cell ($e,l_z,I_3$) 
in equation (\ref{eli_int}) with equally spaced $8\times8\times8$ points, 
and equation (\ref{veli_int}) with equally spaced $128\times128$ points by the rectangle method. 
Here, $\lambda$ and $\nu$ are the positions in the spheroidal coordinate. 
The relation between $(\lambda,\nu)$ and $(R,z)$ are given by 
\begin{equation}
	R^2=\frac{(\lambda-a^2)(\nu-a^2)}{c^2-a^2}, z^2=\frac{(\lambda-c^2)(\nu-c^2)}{a^2-c^2},
\end{equation}
and $V_\mathrm{eff}$ is given by
\begin{equation}
	V_\mathrm{eff}(\tau)=\frac{I_2}{\tau-a^2}+\frac{I_3}{\tau-c^2}-\frac{GM}{c+\sqrt{\tau}}, 
\end{equation}
where $\nu_\mathrm{max}$, $\lambda_\mathrm{min}$ and $\lambda_\mathrm{max}$ 
are the solutions of $E=V_\mathrm{eff}(\tau)$ \citep{dez85}. 
We calculate $\nu_\mathrm{max}$ by the bisection method \citep{pre92}, 
and calculate $\lambda_\mathrm{min}$ and $\lambda_\mathrm{max}$ 
by the golden section search \citep{pre92} and the bisection method. 
As the case of the spherical target model, 
we give the diagnostic quantity $f_\mathrm{dif}$ as
\begin{equation}
	f_\mathrm{dif}=\frac{\sum_{e,l_z,i_3}|w_{el_z i_3}-W_{el_z i_3}|}{\sum_{e,l_z,i_3}W_{el_z i_3}}\times100~\%,
			\label{f_ax_eq}
\end{equation}
where $w_{el_z i_3}$ is the modelling mass weight that 
is calculated by the M2M method on the $(e,l_z,i_3)$ cell.

The second diagnostic quantity is a root mean square (RMS) difference 
between the quantity (mass or intrinsic velocity moments) of the target and the modelling. 
Because the bias of the phase of particles may give the fluctuations for the observables and the RMS values due to the finiteness of the particle number, 
we use multiple phase of observables to calculate the RMS values. 
Therefore, to calculate the RMS values, 
the modelling observables are chosen in every 50 steps from the last 5000 steps. 
The RMS values at each grid $j$ is defined as
\begin{equation}
	\mathrm{RMS}(y_{k,j})=\sqrt{\frac{\sum^{I}_{i=1}(y_{k,j}(t_i)-Y_{k,j})^2}{I}},
	\label{eq_rms}
\end{equation}
where $t_i$ is the time of $i$-th sample, and $I$ is the number of the samples. 
Besides, we give the average of the RMS values 
normalized by the target values for the whole region (averaged RMS) as
\begin{equation}
	\overline{\mathrm{RMS}}(y_k)=\frac{\sum_{j} (\mathrm{RMS}(y_{k,j})/Y_{k,j})}{J}.
\end{equation}
For the RMS of the mass distribution, 
we set that the grids to assess the RMS values 
is same as the grids of the target mass observable. 
On the other hand, for the RMS of the velocity dispersion distribution, 
we use equally spaced grids in the radial 30 shells, 
and the grids extend to $r=r_{\mathrm{max}}$ in the case of the spherical target model. 
In the case of the axisymmetric target model, 
we use equally spaced grids in the meridional ($R-z$) plane 
and the grids extend to $R=z=R_{\mathrm{max}}$ with 32$\times$32 grid points.

Thus, we investigate the degrees of accuracy for the recovery of the spherical and the axisymmetric target models 
by using the two diagnostic quantities $f_\mathrm{dif}$ and $\overline{\mathrm{RMS}}(y_k)$.

%% file: 4result.tex
\section{RESULTS}\label{sec_5}

We show the degree of accuracy for the recovery of the DFs 
for the spherical Plummer model and the axisymmetric st\"{a}ckel model 
to estimate how accurately the DFs (templates) can be constructed. 
To show the results for a fiducial case, 
we fix the parameters such as the particle number, data number, and initial condition in this section. 
Not only the recovery of the DFs but also the recovery of the mass and kinematics are shown 
to comprehend the causes of the errors for the reconstruction of the DFs. 

\subsection{Plummer models}

\subsubsection{Isotropic case}\label{sec_res_plu_iso}

We represent the degree of accuracy for the recovery of the mass, kinematics, and DFs 
for the isotropic ($q=0$) Plummer model. 
\begin{figure}
	\hspace{10mm}
\includegraphics[width=80mm]{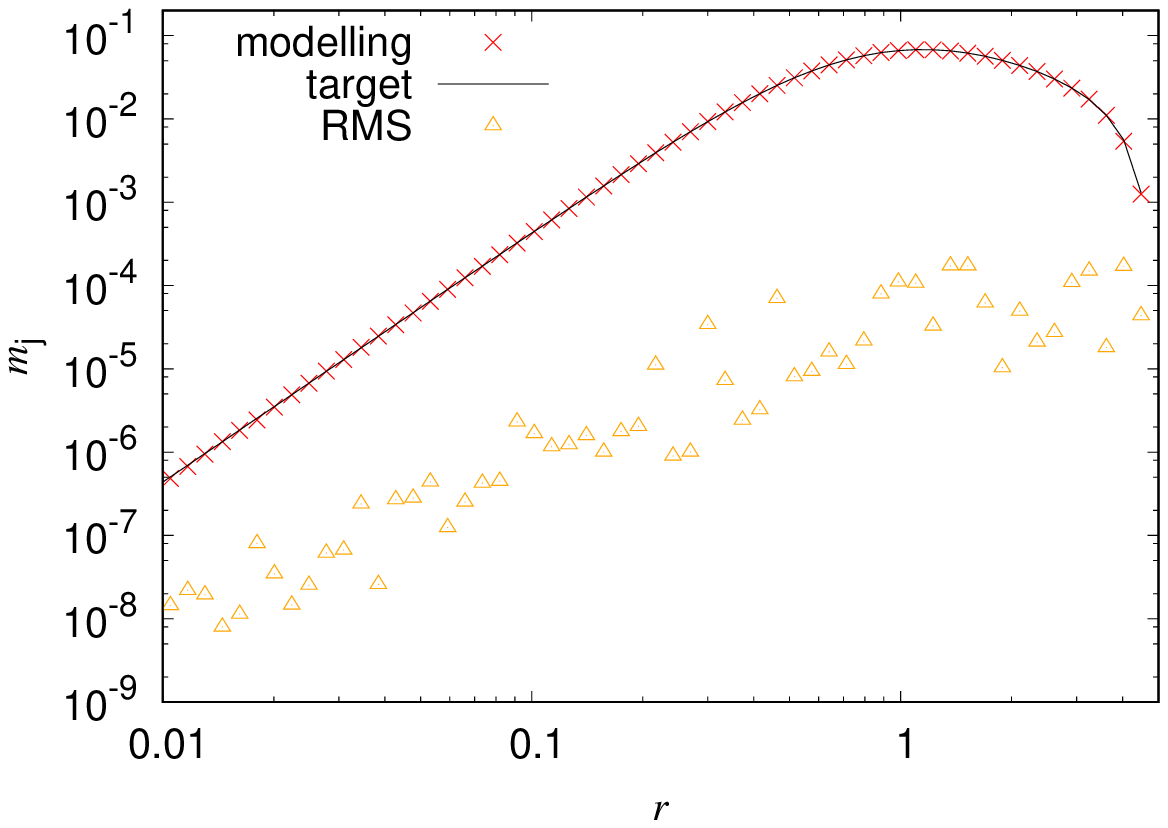}
\includegraphics[width=80mm]{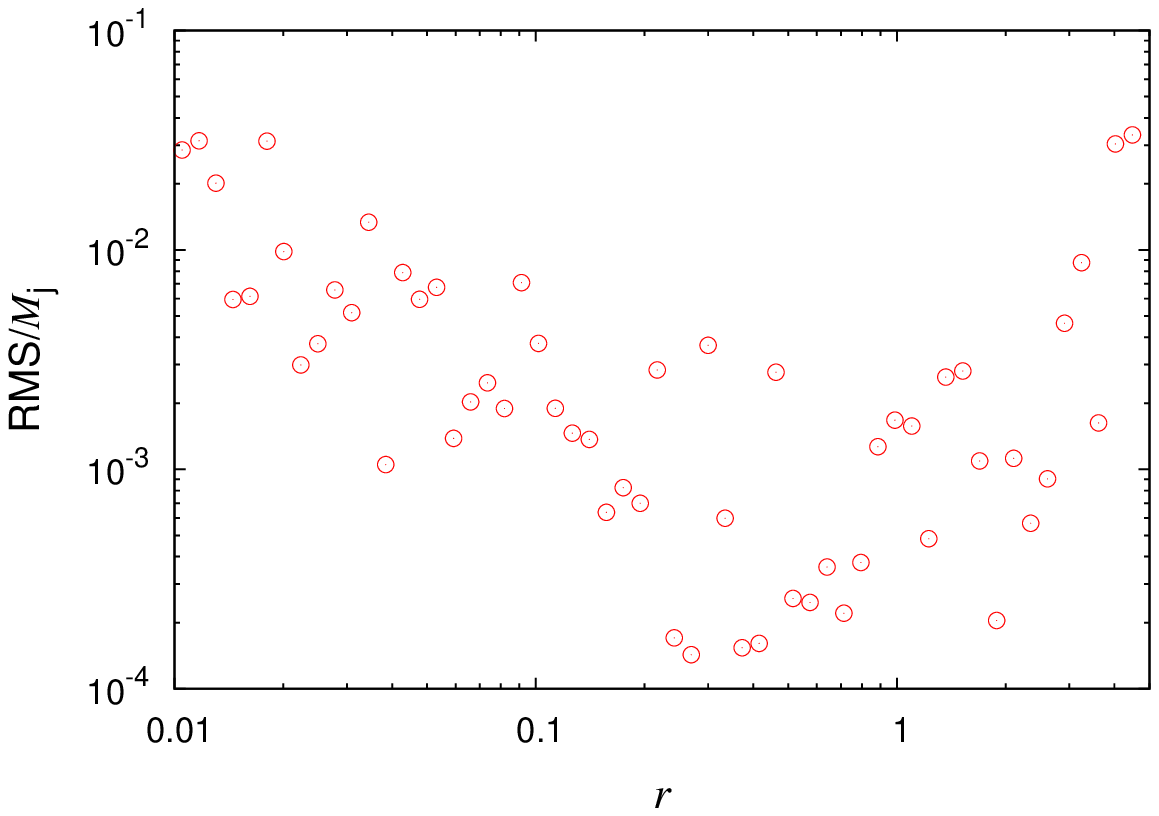}
\caption{
	{\it Left}: Red cross plots, black line and orange triangle plots represent 
	the mass distribution for the modelling, target, and RMS values, respectively. 
	{\it Right}: The mass distribution for the RMS values normalized by the target values. 
	The target model is isotropic ($q=0$) Plummer model.
}
\label{mas_sph}
\end{figure}
First, we verify the degree of accuracy for the recovery of the mass distribution. 
The highly accurate recovery of the mass distribution is important to recover the DFs accurately. 
Because the kinematics are recovered by using the mass weighted quantities ($b_{n,p}=m_{p}h_{n,p}$), 
the errors of the reconstruction for the mass distribution also 
cause the errors of the reconstruction for the kinematics. 
The left panel of Fig. \ref{mas_sph} shows the mass distribution 
for the modelling, target, and RMS values, which are defined in equation (\ref{eq_rms}), in $r\in [0.01,5]$. 
From this panel, the RMS values are about two orders of magnitude lower than the target values. 
The average of the RMS values normalized by the target values $\overline{\mathrm{RMS}}(m)$ is given by 
\begin{equation}
	\overline{\mathrm{RMS}}(m)=\frac{\sum_{j,r_j\in [0.01,5]} (\mathrm{RMS}(m_{j})/M_{j})}{J}=0.36\%.
\end{equation}
This value is consistent with the result of the middle panel of Fig. $11$ in \citet{del07} that 
the recovery of the mass distribution for an isotropic spherical target model has uncertainties of $\sim 1 \%$. 
Thus, the mass distribution is recovered with about equal to or less than one percent error for the isotropic spherical target models. 
The right panel of Fig. \ref{mas_sph} shows the RMS values normalized by the target values.
This panel indicates that the higher errors are seen in the inner and the outer regions. 
In outer regions, the target values are immediately reduced due to the cut off radius of $r_\mathrm{max}=5$. 
Also, in inner regions, the target values are largely reduced because of the functional form of the target model. 
The accurate recovery of the small target values is presumably difficult 
because of the discrete grids of the mass observables or the finite number of the M2M particles. 
The dependence of the degree of accuracy for the recovery of the DFs 
on the particle number and the data number 
is investigated in Section \ref{subsec_par}.

\begin{figure}
	\begin{center}
\includegraphics[width=120mm]{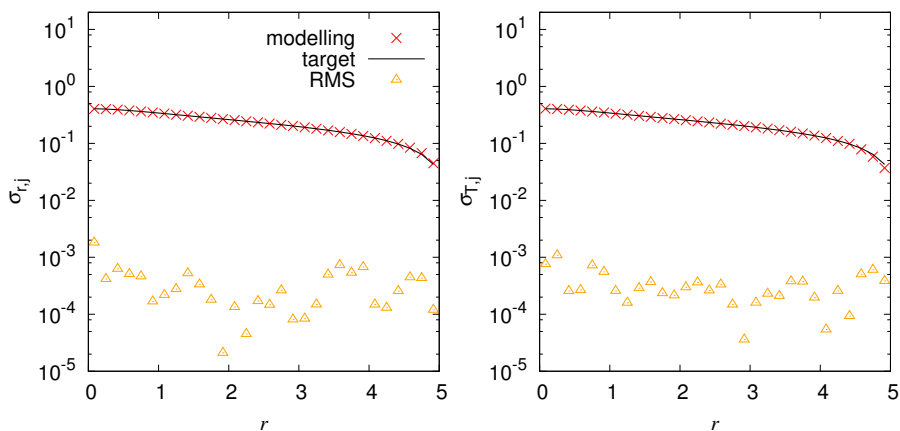}
\caption{
	The radial (left) and tangential (right) velocity dispersion distributions 
	for the modelling (red cross), target (black line), and RMS values (orange triangle). 
	The target model is the isotropic Plummer model. 
}
\label{vel_sph}
\end{center}
\end{figure}

Second, we investigate the degree of accuracy for the recovery of the kinematics. 
The accurate recovery of kinematics is necessary to recover the DFs accurately. 
Fig. \ref{vel_sph} shows the velocity dispersion distribution for the modelling, target and RMS values. 
From this figure, the RMS values in the outer regions ($r\gtrsim4$) are higher than those in the other regions. 
This tendency is similar to the recovery for the mass distribution. 
The errors for the recovery of the mass distribution 
presumably cause the errors for the recovery of the velocity dispersion distribution as mentioned in the previous paragraph. 
Overall, the RMS values are about two orders of magnitude lower than the target values. 
As a result, the averages of the RMS values normalized by the target value 
for the radial ($\overline{\mathrm{RMS}}(\sigma_r)$) and tangential directions ($\overline{\mathrm{RMS}}(\sigma_\mathrm{T})$) 
are $0.81\%$ and $1.04\%$, respectively. 
Hence, the velocity dispersion distribution is recovered with about one percent error for the isotropic spherical target model. 

\begin{figure}
		\hspace{10mm}
	\includegraphics[width=80mm]{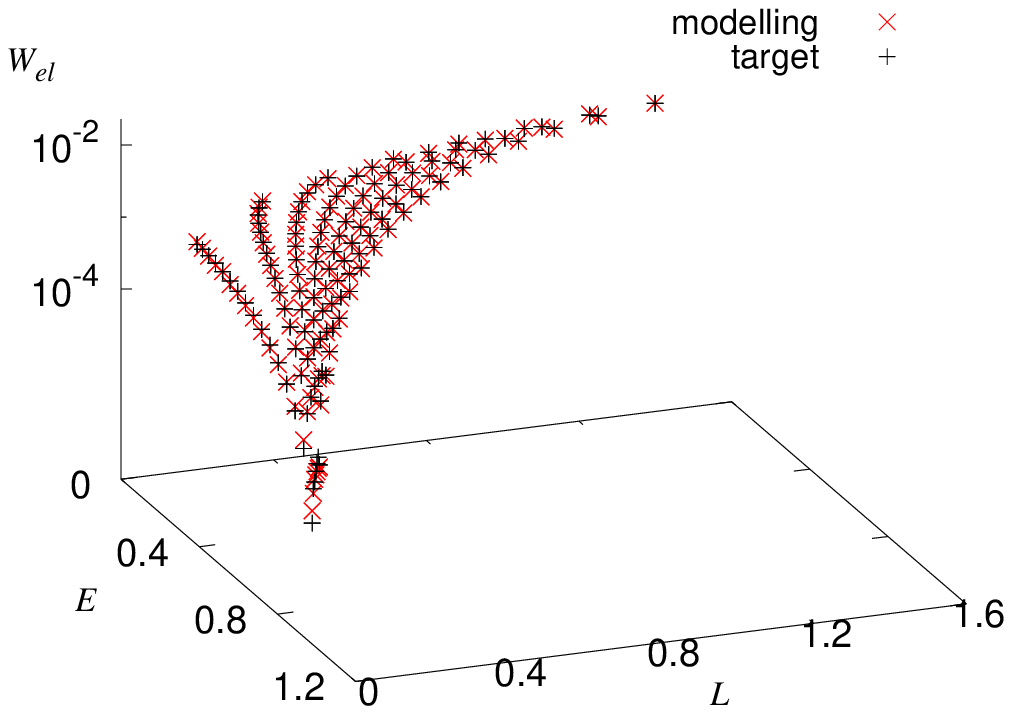}
\hspace{-10mm}
	\includegraphics[width=80mm]{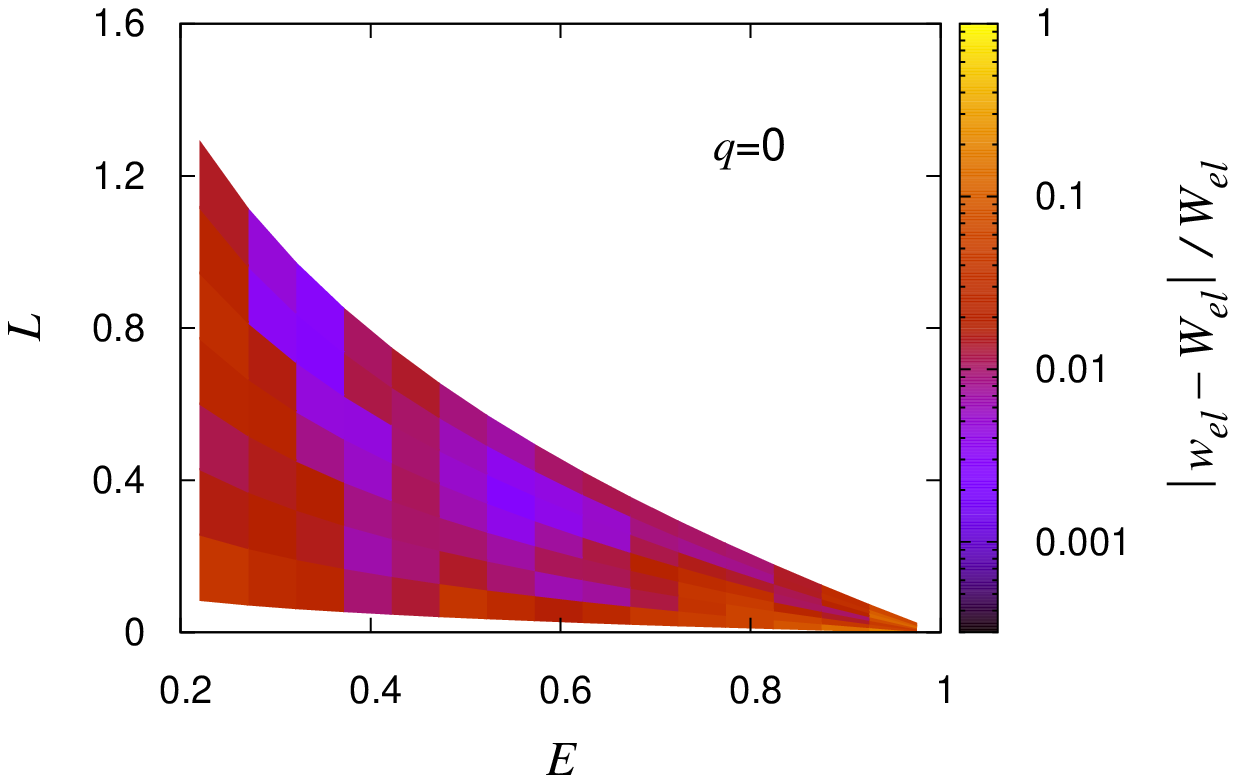}
	\caption{
		Recovery of the DF for the isotropic ($q=0$) Plummer model. 
		{\it Left}: The modelling weight (red cross) and the target weight (black plus) values 
		as functions of the energy ($E$) and the angular momentum ($L$).  
		{\it Right}: The differences between the modelling weight values ($w_{el}$) and the target weight values ($W_{el}$) 
		normalized by the target weight values as functions of the energy and the angular momentum. 
	}
\label{el_sph}
\end{figure}
Next, we indicate the degree of accuracy for the recovery of the DFs. 
The left panel of Fig. \ref{el_sph} shows the distributions of the target (green) and modelling (red) weights 
as functions of the binding energy ($E$) and the total angular momentum ($L$). 
As indicated in this panel, 
the target weight values ($W_{el}$) is low in the high $E$ and low $L$ regions. 
The right panel of Fig. \ref{el_sph} shows 
the normalized differences $|w_{el}-W_{el}|/W_{el}$ in $E,~L$ grids. 
From this panel, 
the high errors are seen in the low $L$, low $E$, and high $E$ regions. 
The high errors in the low $L$ or high $E$ regions are supposed to be due to the low values for $W_{el}$ in these regions. 
The accurate recovery of the small weight regions will be difficult 
as mentioned in the recovery of the mass distribution. 
On the other hand, the errors in the low $E$ regions presumably 
relate to the high errors of the mass distribution in the outer (large $r$) regions. 
This is because the particles that have low $E$ are often in the outer regions. 
On the whole, the differences between the modelling and target values 
are almost two orders of magnitude lower than the target values. 
As a result, the degree of accuracy for the recovery of the DF $f_\mathrm{dif}$ 
represented in equation (\ref{f_sph_eq}) is $1.55\%$. 
Thus, the DF (template) for the isotropic spherical target model is recovered with about one percent error. 


\subsubsection{Anisotropic case}\label{sec_ani}

We investigate the degree of accuracy 
for the recovery of the anisotropic spherical target models shown in Section 3.1 and Appendix A. 
We found that 
for the anisotropic spherical target models, 
$\overline{\mathrm{RMS}}(m)$ for the radially anisotropic $(q=0.5)$ 
and the tangentially anisotropic $(q=-0.5)$ models 
are $0.21~\%$ and $0.13~\%$, respectively. 
Thus, 
$\overline{\mathrm{RMS}}(m)$ for the anisotropic models are 
as low as that for the isotropic target model with $\overline{\mathrm{RMS}}(m) =0.36\%$. 
This result implies that the errors for the reconstruction of the mass distribution 
are not caused by the anisotropy of the target models. 

\begin{figure}
	\begin{center}
\includegraphics[width=120mm]{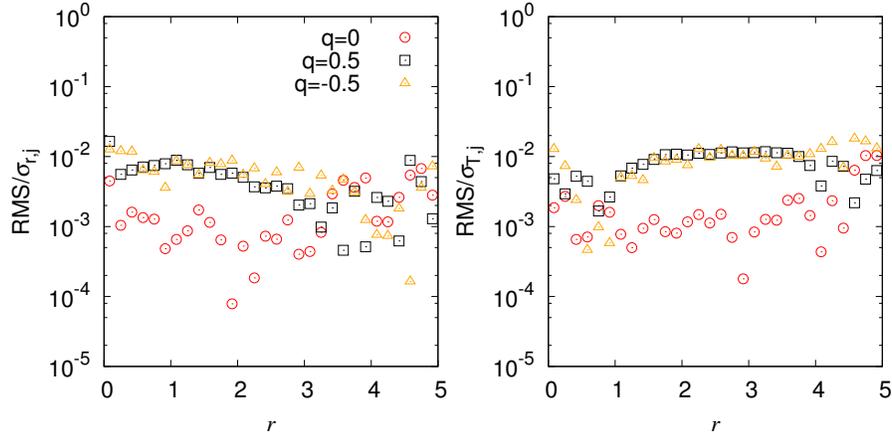}
\caption{
	The RMS values normalized by the target values 
	for the radial (left) and tangential (right) velocity dispersion distributions. 
	The red circle, black square, and orange triangle plots 
	represent the results of $q=0$, $q=0.5$, and $q=-0.5$, respectively. 
}
\label{vel_sph_065}
\end{center}
\end{figure}
Fig. \ref{vel_sph_065} shows the RMS values normalized by the target values 
for the radial (left) and tangential (right) velocity dispersion distributions. 
This figure indicates that the normalized RMS values 
for the isotropic model (red circle) is lower than that for the other models. 
Meanwhile, the RMS values in $r=4-5$ are more dispersed than that in the other regions. 
These high RMS values are presumably caused by the high RMS values 
for the mass distribution in these regions as mentioned in the case of the isotropic spherical target model. 
Overall, 
$\overline{\mathrm{RMS}}(\sigma_r)$ and $\overline{\mathrm{RMS}}(\sigma_\mathrm{T})$ 
are $1.31\%$ and $2.33\%$ for $q=0.5$, 
and are $1.91\%$ and $2.84\%$ for $q=-0.5$, respectively. 
These values are almost consistent with the results of \citet{mor12} 
that the recovery of the velocity dispersion distribution 
for spherical anisotropic models has uncertainties of about a few percent. 

\begin{figure}
\hspace{5mm}
\includegraphics[width=85mm]{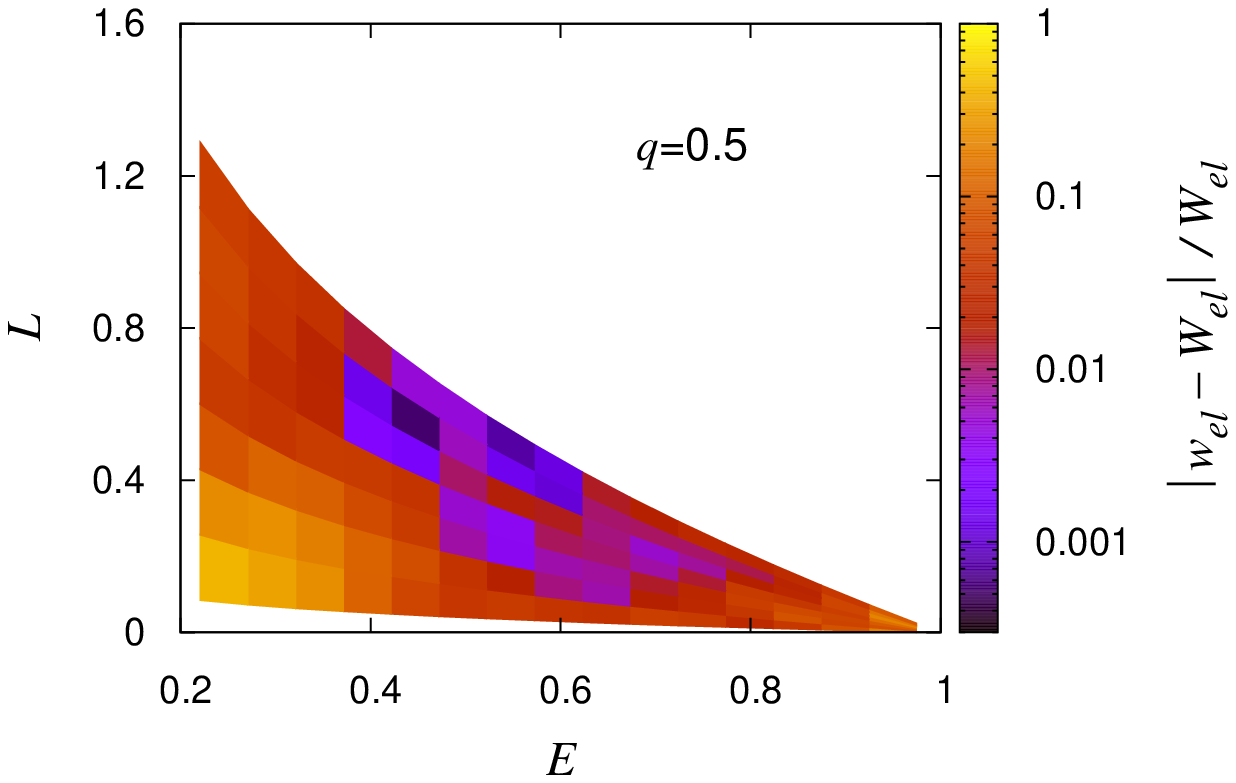}
\hspace{-5mm}
\includegraphics[width=85mm]{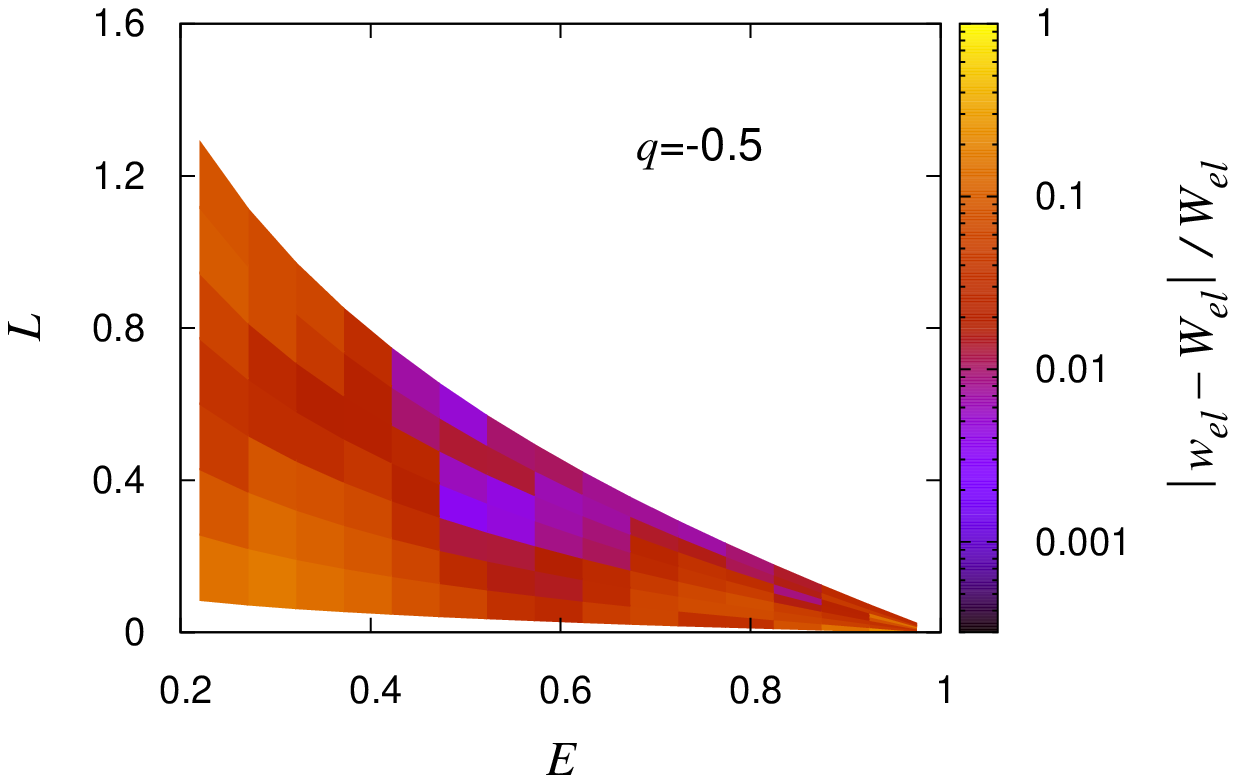}
\caption{
	Same as right panel of Fig. \ref{el_sph}, 
	but left and right panels are the results for 
	the radially ($q=0.5$) and tangentially ($q=-0.5$) anisotropic target models, respectively. 
} 
\label{el_sph_065}
\end{figure}
Fig. \ref{el_sph_065} shows 
the normalized differences $|w_{el}-W_{el}|/W_{el}$ 
for the radially anisotropic ($q=0.5$, left panel) 
and tangentially anisotropic ($q=-0.5$, right panel) target models. 
The tendency of the distributions of the errors for these anisotropic models 
is similar to that for the isotropic model. 
On the other hand, the values of the errors for the anisotropic models 
are a few times larger than that for the isotropic model. 
As a result, 
$f_\mathrm{dif}$ 
for the radially and tangentially anisotropic models are $3.61$ and $3.38\%$. 
Hence, the degree of accuracy for the recovery of the anisotropic DFs 
is about two times larger than that of the isotropic DF. 
The higher errors for the recovery of the anisotropic DFs 
are presumably related to the higher errors for the recovery of the kinematics. 
In fact, the errors for the recovery of the velocity dispersion distribution for the anisotropic target models 
are also about two times larger than those for the isotropic target model. 
Consequently, the DFs are typically recovered 
with a few percent errors in the cases of the spherical target models. 
Hence, if the target galaxy is a spherical symmetry, 
the templates (DFs) can be typically constructed with a few percent error. 


\subsection{Axisymmetric models}

In this section, we show the accuracies for the recovery of 
the mass distribution, the velocity dispersion distribution and the DF 
for the axisymmetric three integral target model (shown in 3.2 and Appendix B). 
\begin{figure}
	\hspace{10mm}
\includegraphics[width=80mm]{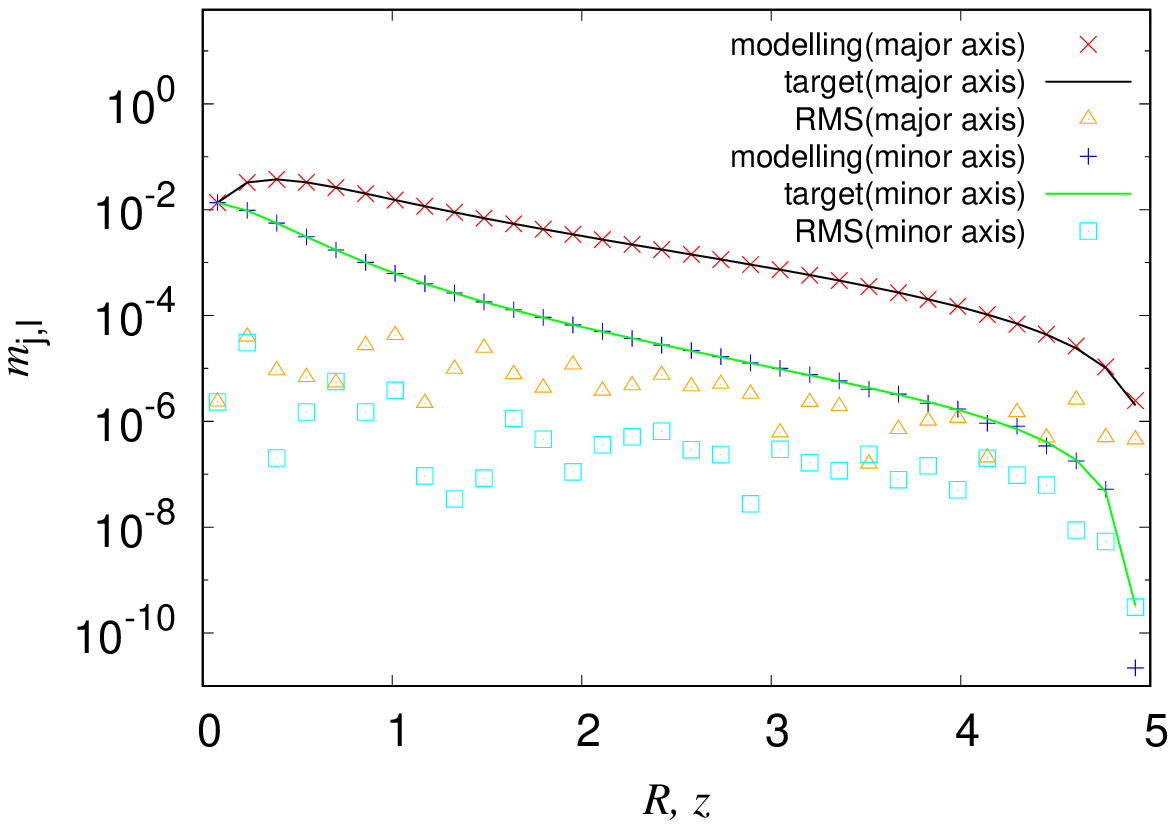}
\includegraphics[width=80mm]{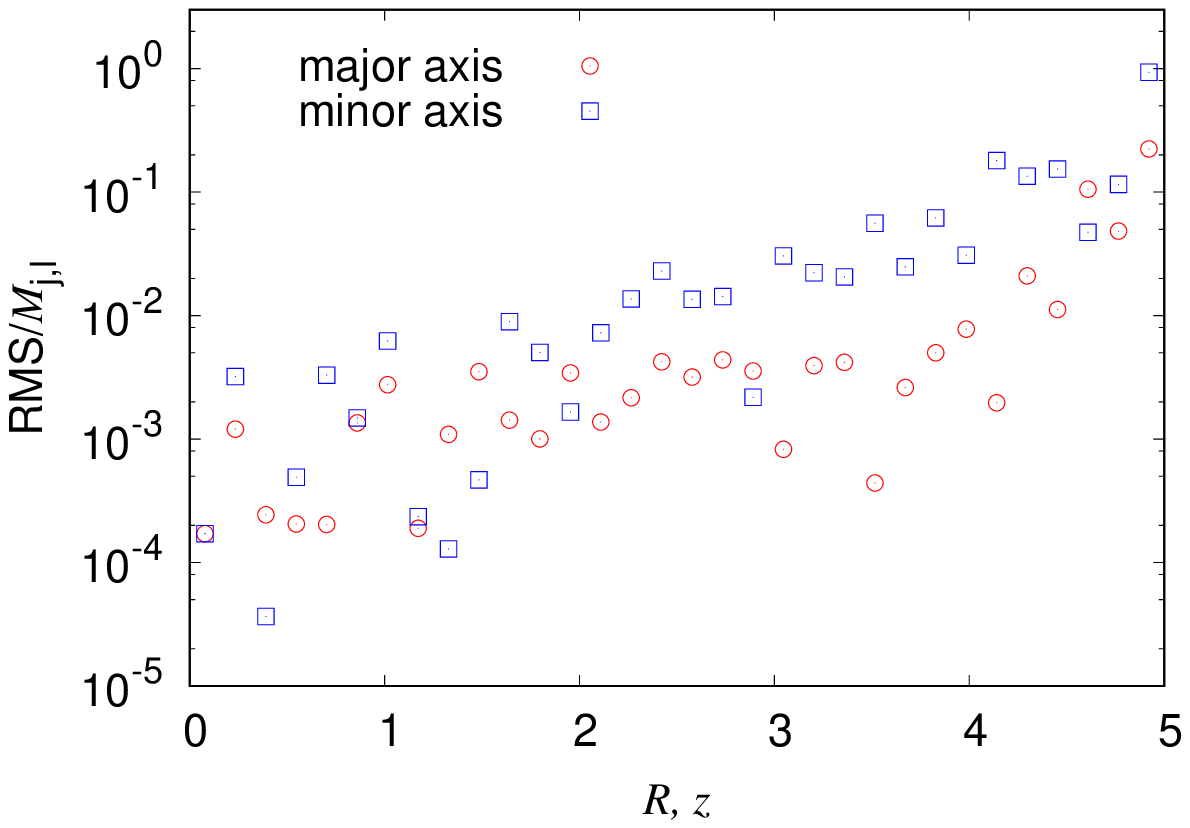}
\caption{
	Recovery of the mass distribution for axisymmetric target model (a) of equations (\ref{ax_model}). 
	{\it Left: }
	Red cross (blue plus), black (green) line, and orange triangle (cyan square) plots 
	represent the modelling, target, and RMS values along the major (minor) axis, respectively. 
	{\it Right: }The RMS values normalized by the target values for the mass distribution
	along the major (red circle) and the minor (blue square) axis. 
}
\label{mas_ax}
\end{figure}
The left panel of Fig. \ref{mas_ax} shows the modelling, target, and RMS values 
of the mass distribution for axisymmetric target model (a) 
along the major axis ($l=0$ in equation (\ref{eq_mass4})) and the minor axis ($j=0$ in equation (\ref{eq_mass5})). 
This panel indicates that the RMS values are about two orders of magnitude lower than the target values. 
In the right panel of Fig. \ref{mas_ax}, 
the RMS values normalized by the target values 
for the mass distribution along the major axis and the minor axis are shown. 
The normalized RMS values are high in the outer regions ($R\gtrsim4$, or $z\gtrsim3$) where the target values are low. 
The tendency that the accurate recovery is difficult in the regions where target values are low 
is also seen in the result for the recovery of the mass distribution and DF for the spherical target models. 
We suppose that this tendency is the fundamental characteristic for the M2M method. 
Consequently, the average of the mass RMS values normalized by the target values 
is 
\begin{equation}
	\overline{\mathrm{RMS}}(m)=\frac{\sum_{j,l} (\mathrm{RMS}(m_{j,l})/M_{j,l})}{J}=1.17\%.
\end{equation}
This value is consistent with the results of Fig. $13$ in \citet{del07} that 
the recovery of the mass distribution for an axisymmetric target model has uncertainties of about a few percent. 

\begin{figure}
	\begin{center}
\includegraphics[width=160mm]{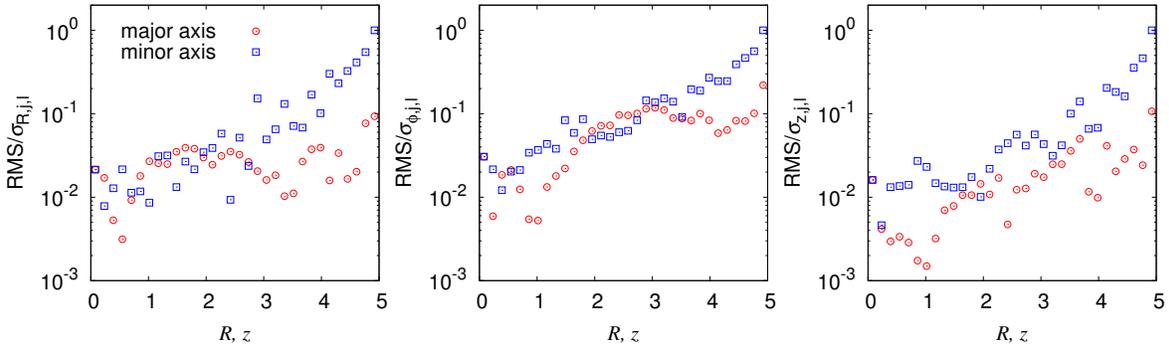}
\caption{
	The radial (left), azimuthal (middle), and $z$-axis (right) velocity dispersion distributions for 
	the RMS values normalized by the target values. 
	Red circle and blue square plots represent the results along the major axis and the minor axis. 
}
\label{vel_ax}
\end{center}
\end{figure}

Fig. \ref{vel_ax} shows the RMS values normalized by the target values 
of the velocity dispersion distribution for target model (a). 
This figure shows that 
the normalized RMS values are high in the large $R$ and $z$ regions. 
The distribution of the high errors for the recovery of the velocity dispersion is similar to that of the mass 
as seen in the right panel of Fig. \ref{mas_ax} and \ref{vel_ax}. 
Therefore, we suppose that the errors of the velocity dispersion distribution are affected by the errors of the mass distribution. 
The relation for the distribution of the high normalized RMS values between the mass and the velocity dispersion distributions 
is also observed in the results for the spherical target model. 
Such a relation is comprehensible because 
the kinematics are recovered by using the mass weighted quantities ($b_{n,p}=m_{p}h_{n,p}$). 
The averages of the normalized  RMS values of the radial ($R$), azimuthal ($\phi$), and $z$-axis velocity dispersions 
$\overline{\mathrm{RMS}}(\sigma_R),~\overline{\mathrm{RMS}}(\sigma_\phi)$, 
and $\overline{\mathrm{RMS}}(\sigma_{z})$ for target model (a) are $3.74,~6.76$, and $3.29\%$, respectively. 
The errors for the recovery of the azimuthal velocity dispersion are higher than the others. 
As also seen from the results along the major axis (red circle) in Fig. \ref{vel_ax}, 
the normalized RMS values for the azimuthal velocity dispersion (middle panel) are higher than that for the others. 
We suppose that these high errors for the recovery of the azimuthal velocity dispersion distribution 
are due to the choice of the initial condition. 
Because the initial condition is the isotropic velocity distribution, 
the azimuthally anisotropic particles are assumed to be deficient.  

\begin{figure}
	\begin{center}
	\includegraphics[width=85mm]{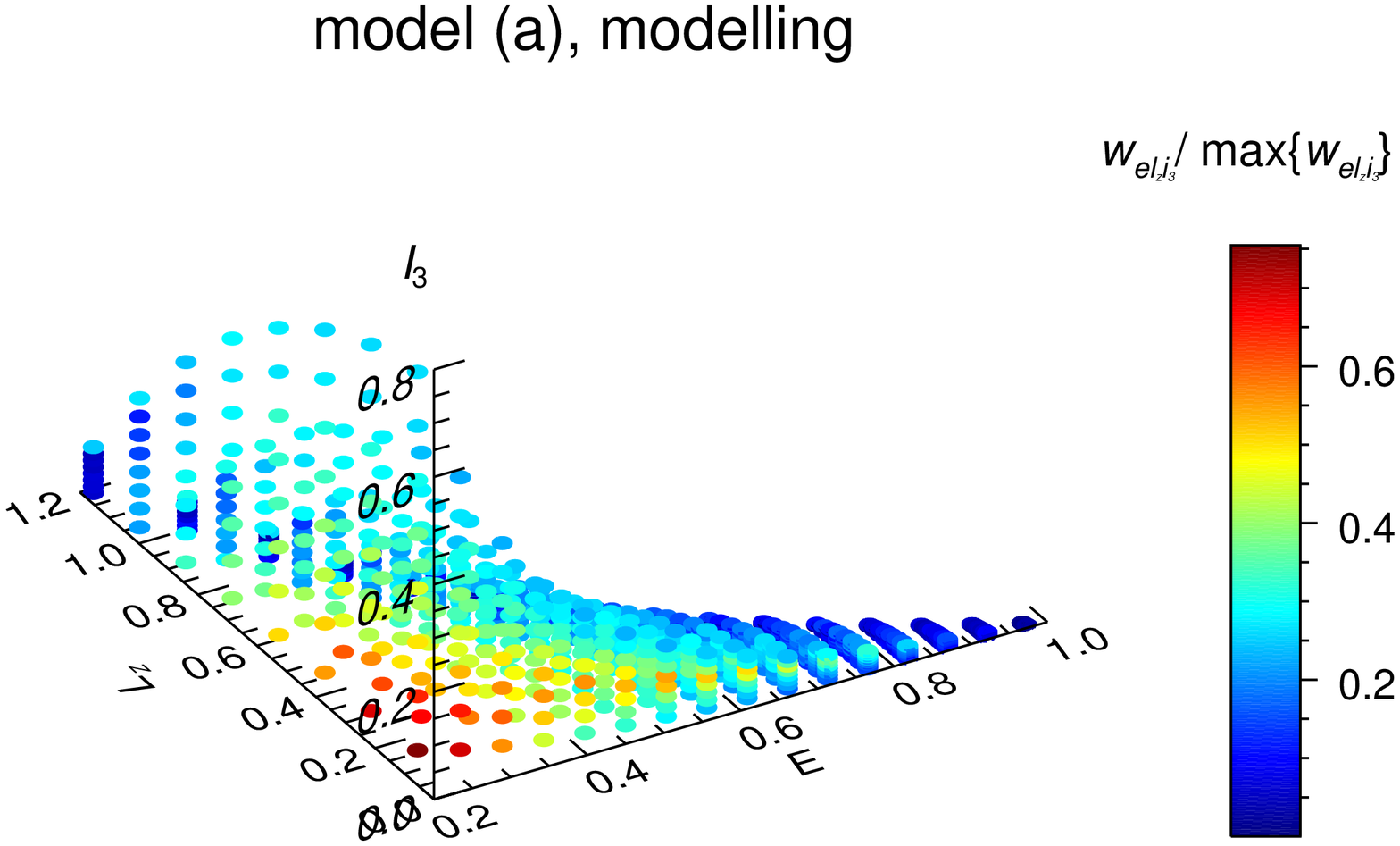}
	\includegraphics[width=85mm]{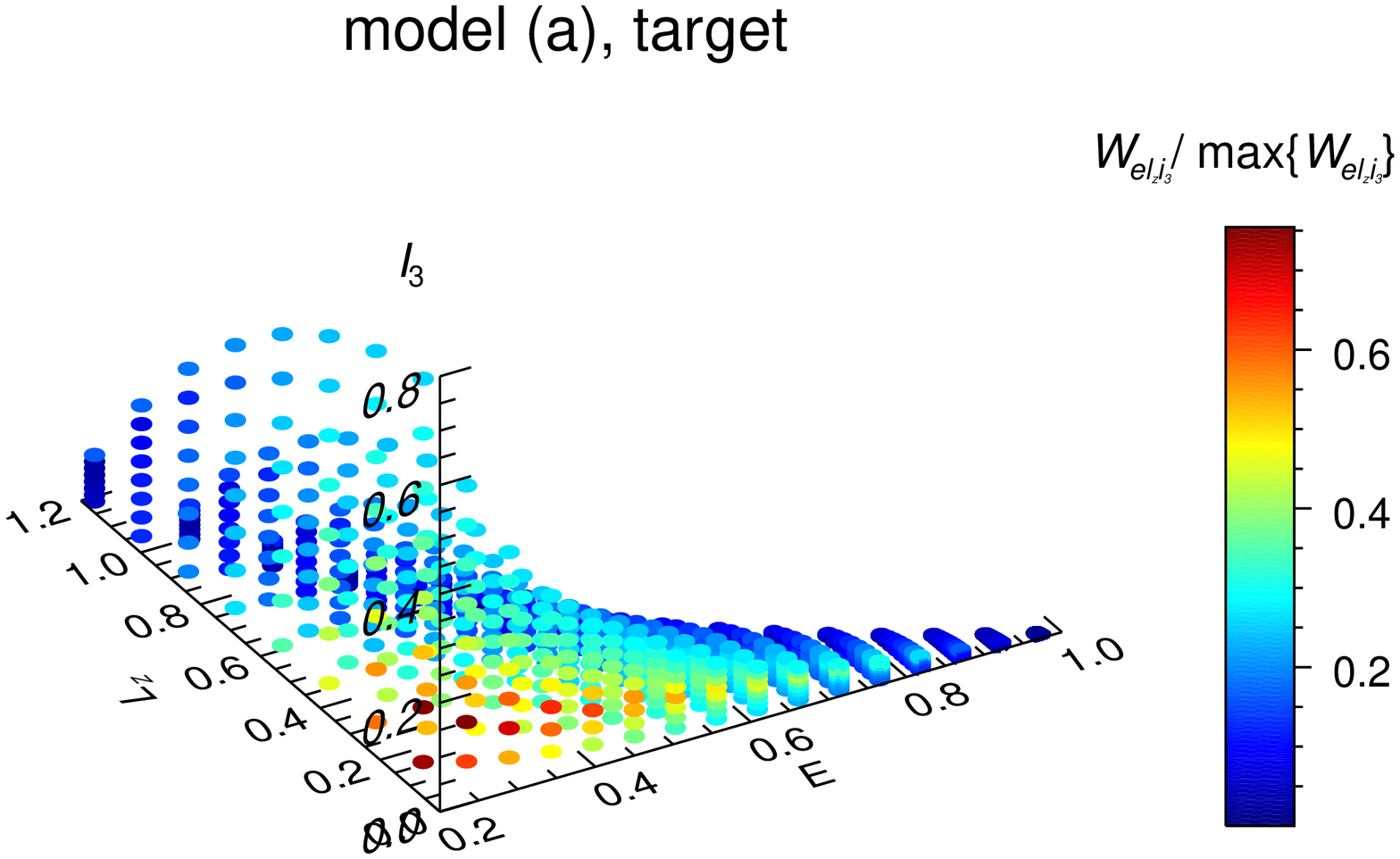}
	\includegraphics[width=85mm]{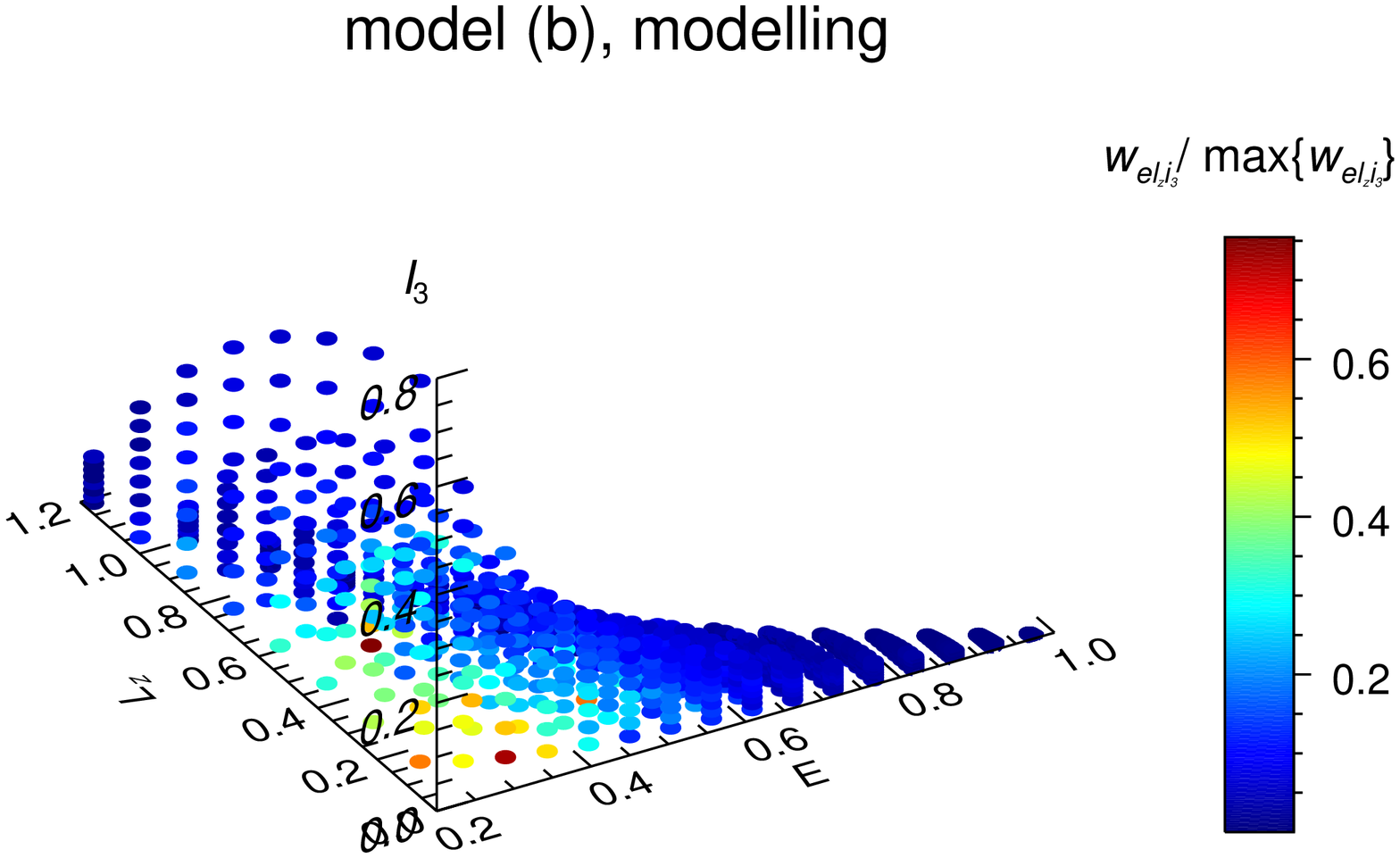}
	\includegraphics[width=85mm]{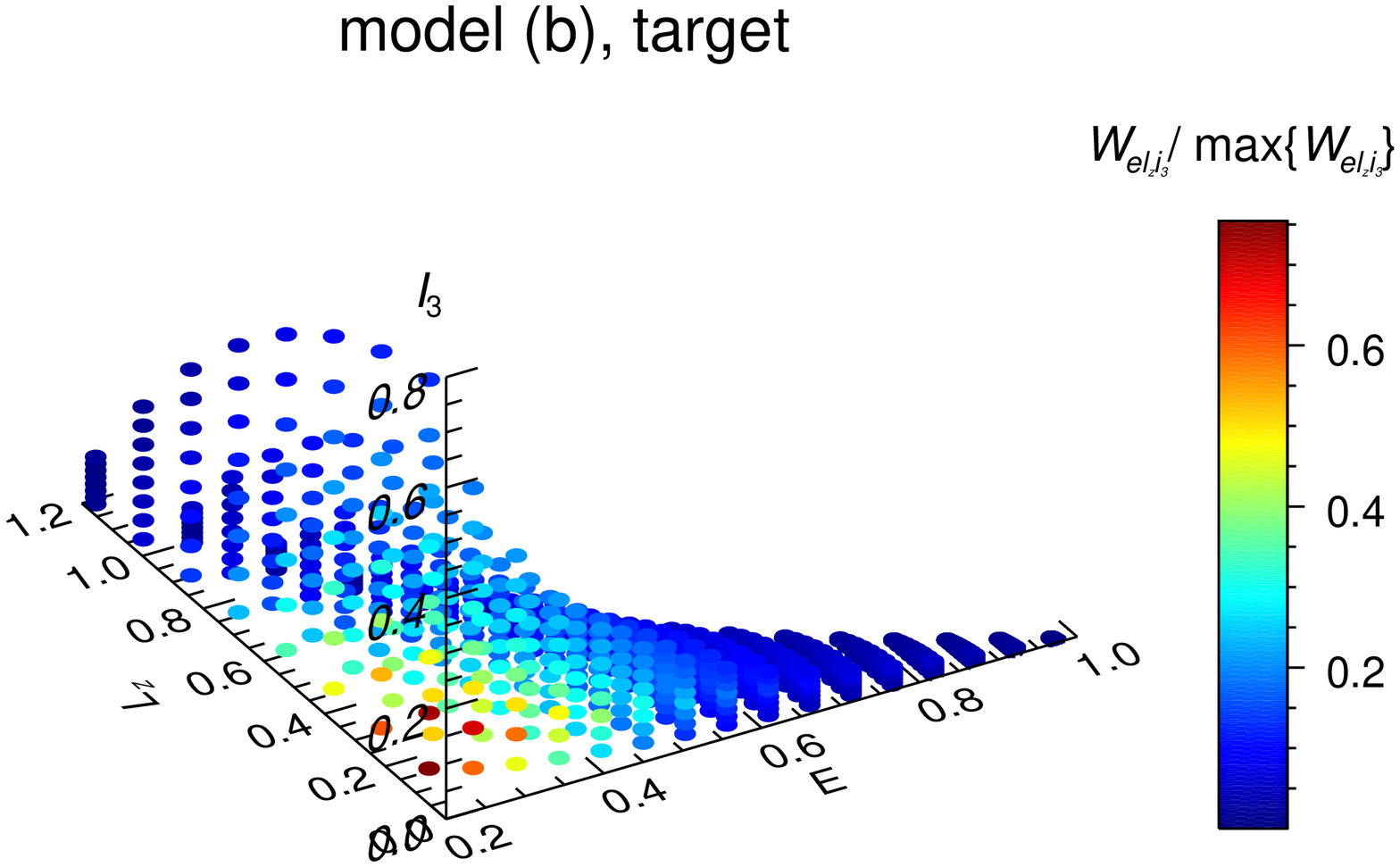}				
	\caption{The DFs as functions of the energy ($E$), the angular momentum ($L_z$), 
		and the third integrals of motion ($I_3$). 
		Left and right panels represent the modelling weight values $w_{el_zi_3}$ 
		and the target weight values $W_{el_zi_3}$. 
		Upper and lower panels are results for target models (a) and (b) of equations (\ref{ax_model}). 
		The color shows the weight values normalized by the maximum weight value among each integral of motion space. 
	}
	\label{iso_el}
	\end{center}
\end{figure}
\begin{figure}
	\begin{center}
	\includegraphics[width=70mm]{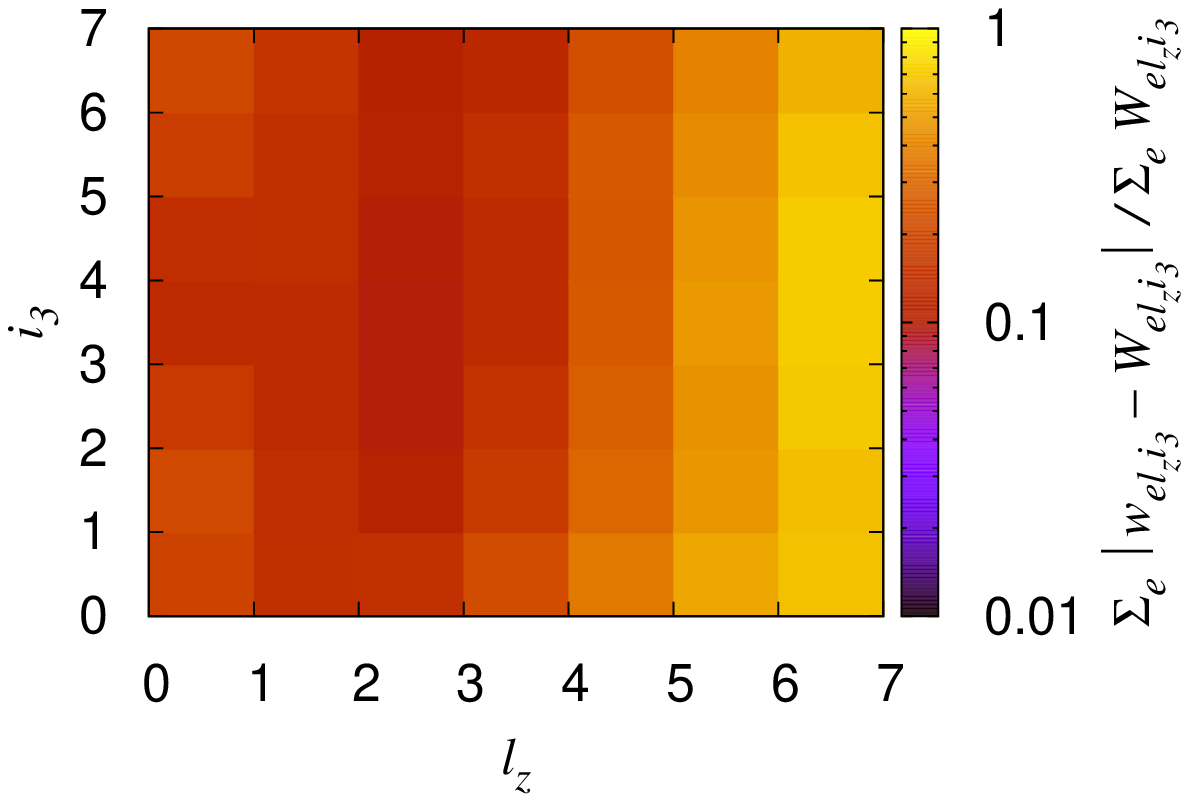}
	\includegraphics[width=70mm]{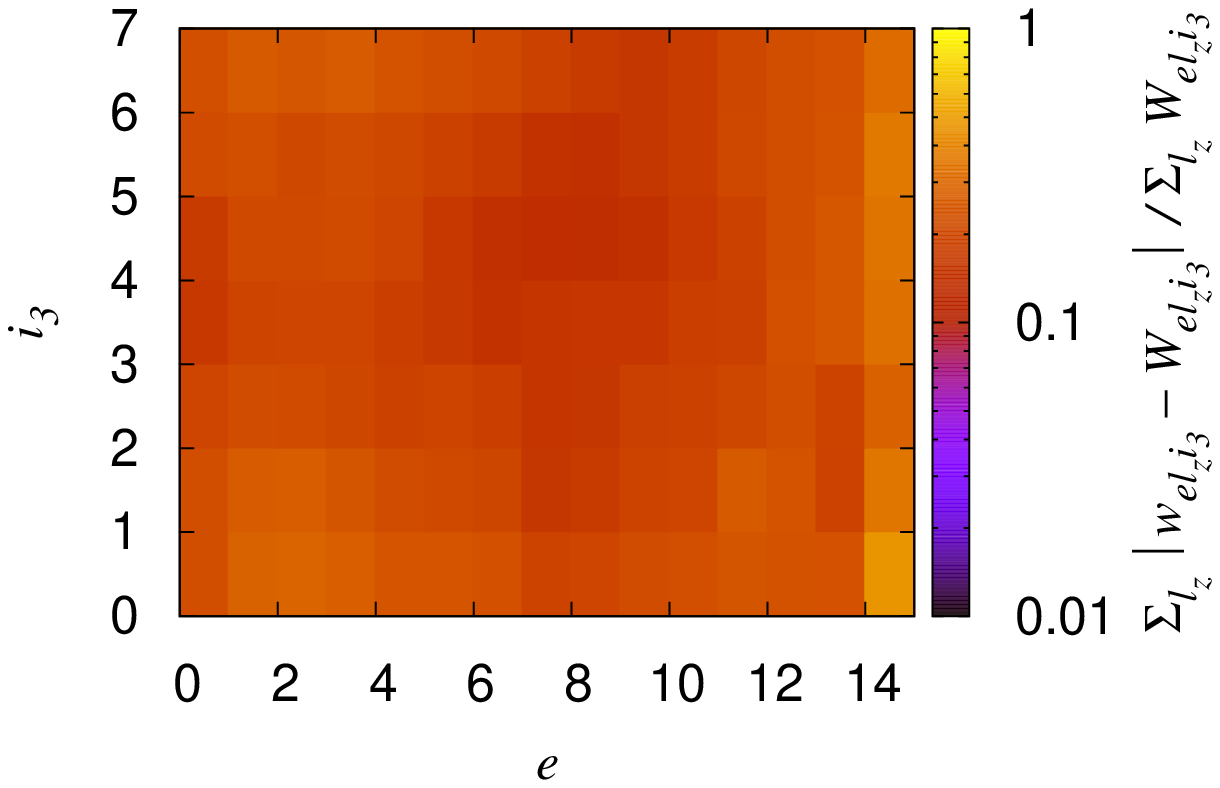}
	\includegraphics[width=70mm]{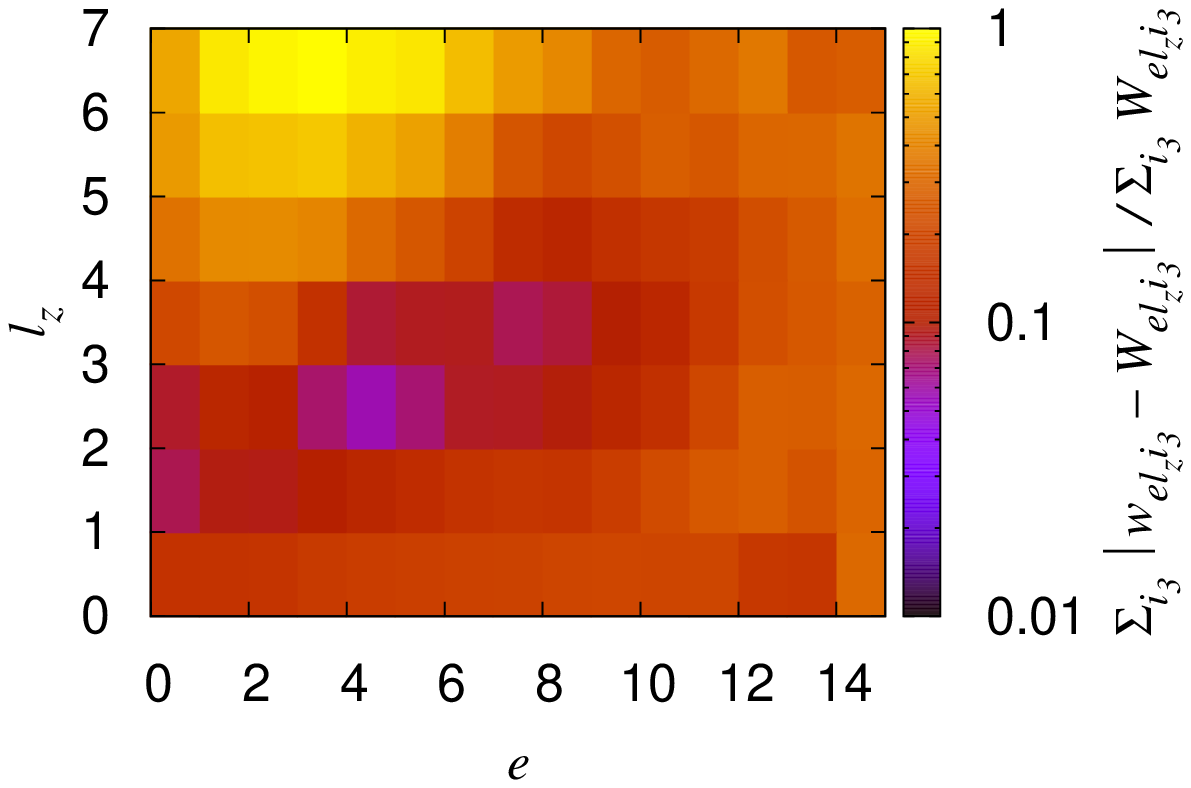}
	\caption{
		The upper left, upper right and lower panels are 
		the sum for the energy ($E$), the angular momentum ($L_z$) and 
		third integral of motion ($I_3$) 
		of $|W_{el_zi_3}-w_{el_zi_3}|$ normalized by the sum of $W_{el_zi_3}$ 
		for target model (a) of equations (\ref{ax_model}), respectively. 
		The lower values of the grid numbers $e,~l_z,~i_3$ correspond to the lower values of $E,~L_z,~I_3$. 
	}
	\label{4_el}
	\end{center}
\end{figure}

Fig. \ref{iso_el} represents the DFs for the modelling (left panels) and the target (right panels) 
for target models (a) (upper panels) and (b) (lower panels). 
The clear difference of the DFs between target models (a) and (b) 
is whether the weights are distributed until high $E$ or not. 
From this figure, the DFs that depend on the three integrals of motion are well recovered 
by the M2M method as shown by the orbit-based method \citep*{cha08,ven08}. 
Fig. \ref{4_el} indicates the normalized errors of the DF for each integrals of motion space. 
($\Sigma|w_{el_zi_3}-W_{el_zi_3}|/\Sigma W_{el_zi_3}$). 
From these panels, the high errors are seen in the high $L_z$ regions. 
On the other hand, in the spherical systems, 
high errors are seen in the low $L$ regions. 
Here, the DF of the axisymmetric target model have large weights in the low $L_z$ regions 
while the DF of the isotropic spherical target model have large weights in the high $L$ regions. 
Therefore, these differences of the traits for the DFs possibly cause the difference of the traits for the distribution of high errors. 
We suppose that the recovery of low weight value regions 
easily contain high uncertainties 
as seen in the recovery of the mass distributions for the spherical and axisymmetric target models and the DFs for the spherical target model. 
From our result, 
$f_\mathrm{dif}$ 
for axisymmetric target model (a) is $19.9~\%$. 
Thus, $f_\mathrm{dif}$ for the axisymmetric target model is significantly larger than that for the spherical target model. 
Such large value of $f_\mathrm{dif}$ for the axisymmetric target model is mostly due to 
the increase of the number of the integrals of motion that are required to represent the DFs. 
The value of $f_\mathrm{dif}$ that depends on the three integral of motion is 
about ten times larger than the value of $f_\mathrm{dif}$ that depends on the two integrals of motion 
even if the errors for the recovery of the velocity dispersion distribution for the model that depends on the three integrals of motion 
is same as that on the two integrals of motion. 
This value of $f_\mathrm{dif}=19.9~\%$ 
is almost consistent with the result of \citet{ven08} 
that the recovery of the DF for the axisymmetric three integrals target model 
by the orbit-based method have uncertainties of $\sim30~\%$. 
These results represent that the DFs (templates) for axisymmetric three integral target models 
are typically recovered with a few tens percent. 
On the other hand, since $f_\mathrm{dif}$ varies according to some parameters, 
we investigate the dependence of $f_\mathrm{dif}$ on several parameters in the next section. 

%% file: 5discussion.tex
\section{DISCUSSION}\label{sec_6}

In this section, we investigate the dependence of 
the degree of accuracy for the recovery of the DFs ($f_\mathrm{dif}$) on some parameters, 
which are the particle number, the data number, the initial condition, 
the higher order velocity moments, 
the entropy parameter, 
and the configurations for the grids of the kinematic observable. 

\subsection{Dependence of the particle number and the data number}\label{subsec_par}

We first investigate the dependence of 
$f_\mathrm{dif}$ on the particle number ($N$) and the data number for several target models. 
The initial condition used in Section \ref{subsec_par} is the Hernquist with Gaussian (same as Section \ref{sec_5}). 
In this initial condition, 
the spatial distribution is given by Hernquist mass model, 
and the velocity distribution is given by the Gaussian distribution 
whose dispersion is given from solving the Jeans equations for the Hernquist potential. 

\subsubsection{Isotropic models}\label{sec_dis_dep_iso}

\begin{figure}
\hspace{10mm}
\includegraphics[width=85mm]{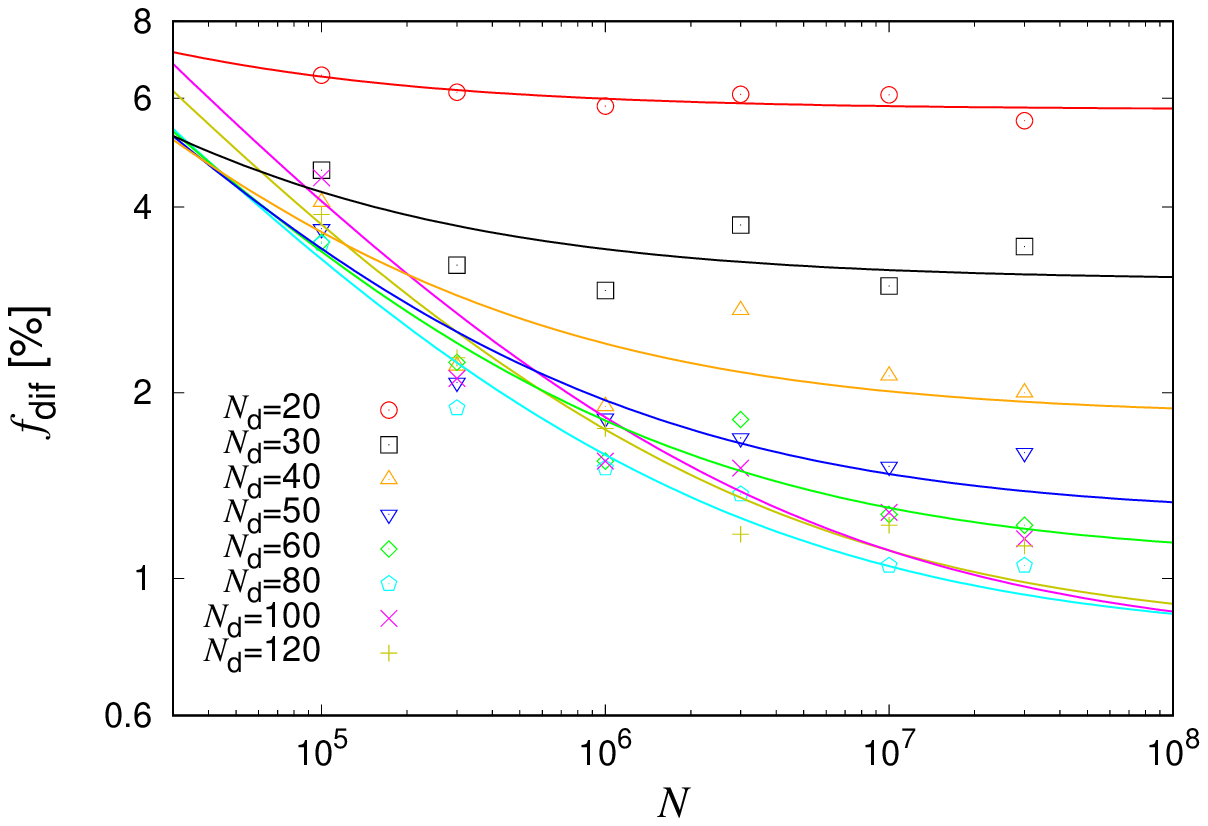}
\includegraphics[width=85mm]{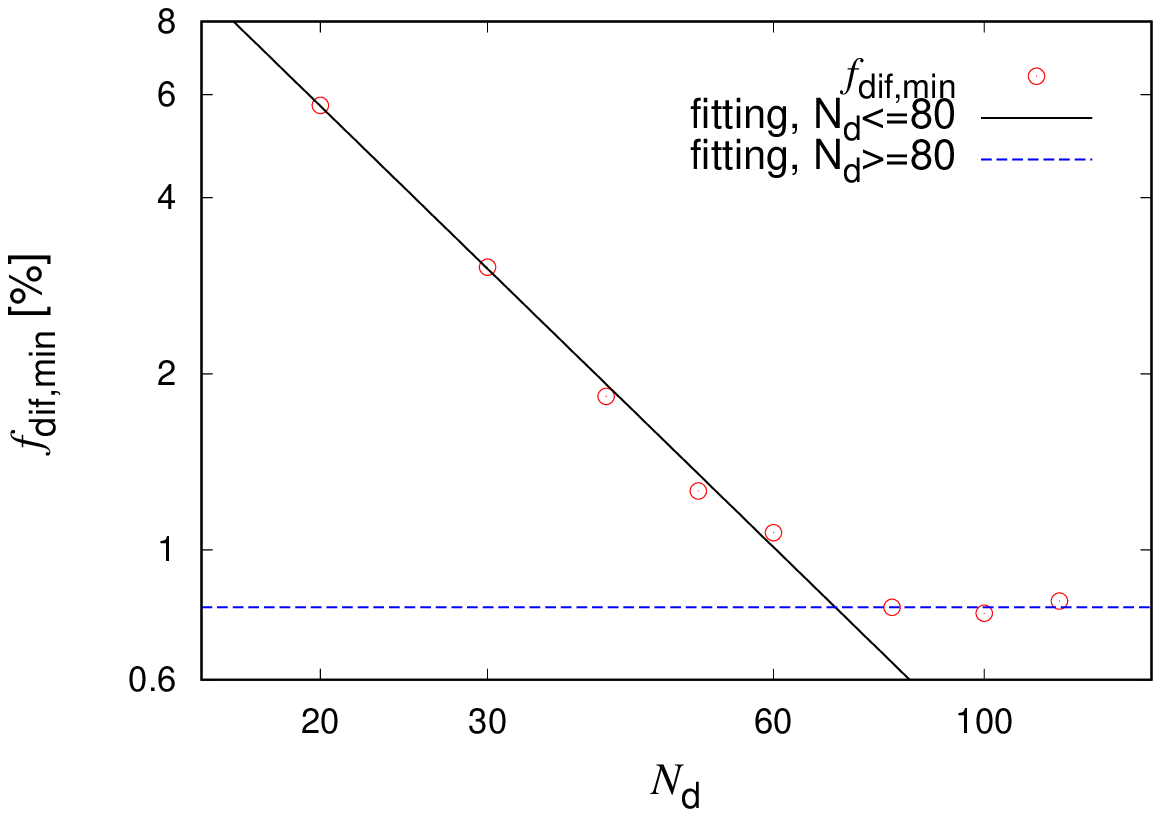}
\caption{
	The left panel shows the degree of accuracy for the recovery of the DFs ($f_\mathrm{dif}$) 
	for the isotropic Plummer model ($q=0$) as a function of the particle number ($N$). 
	Red circle, black square, orange triangle, blue inverted triangle, green diamond, 
	cyan pentagon, magenta cross, and yellow plus plots indicate 
	the results for $N_\mathrm{d}=$20, 30, 40, 50, 60, 80, 100, and 120, respectively. 
	Each line represents the curve fitted by $f_\mathrm{dif}=a_0\times N^{-0.5}+f_\mathrm{dif,min}$ for concolorous plots. 
	The right panel shows $f_\mathrm{dif,min}$ as a function of the data number $N_\mathrm{d}$. 
	The lines represent the curves fitted by $f_\mathrm{dif,min}=a_0\times N_\mathrm{d}^{b_0}$ (black line) in $N_\mathrm{d}\leq80$ 
	and $f_\mathrm{dif,min}=a_0$ (blue dashed line) in $N_\mathrm{d}\ge80$. 
}
\label{df_nf_sph}
\end{figure}

To investigate the dependence of $f_\mathrm{dif}$ on $N$ and the data number ($N_\mathrm{d}$), 
we reconstructed the DF for the isotropic Plummer target model 
using several values of $N$ and $N_\mathrm{d}$. 
The left panel of Fig. \ref{df_nf_sph} shows $f_\mathrm{dif}$ 
as a function of $N$ for several $N_\mathrm{d}$. 
Each line is fitted by a function $f_\mathrm{dif}=a_0\times N^{-0.5}+f_\mathrm{dif,min}$ for each $N_\mathrm{d}$, 
where $a_0$ and $f_\mathrm{dif,min}$ are fitting parameters. 
Here the power law index of the fitting curve is given by $-0.5$ 
because we suppose that the errors caused by the shortage of $N$ 
shows behavior similar to the Poisson noise. 
As can be seen from the left panel of Fig. \ref{df_nf_sph}, 
the plots are almost well fitted by the function $f_\mathrm{dif}=a_0\times N^{-0.5}+f_\mathrm{dif,min}$. 
The fitted lines indicate that 
the convergences of $f_\mathrm{dif}$ to the minimum values ($f_\mathrm{dif,min}$) are in $N \sim 10^6-10^8$. 
Furthermore, as $N_\mathrm{d}$ increases, 
then $N$ that is required to achieve $f_\mathrm{dif} \sim f_\mathrm{dif,min}$ increases. 
From these results, 
we suggest that the particle number that is required to recover the DFs with high accuracies 
is larger than the particle number used by the previous studies for the M2M method ($\sim 10^6$) 
especially for the large data number. 

The right panel of Fig. \ref{df_nf_sph} shows 
the dependence of $f_\mathrm{dif,min}$ on $N_\mathrm{d}$, 
where $f_\mathrm{dif,min}$ is the fitting parameter in the left panel of Fig. \ref{df_nf_sph}. 
The value of $f_\mathrm{dif,min}$ represents $f_\mathrm{dif}$ for a sufficiently large $N$. 
To elucidate the dependence of $f_\mathrm{dif,min}$ on $N_\mathrm{d}$, 
we fit $f_\mathrm{dif,min}$. 
Since the dependence of $f_\mathrm{dif,min}$ on $N_\mathrm{d}$ is abruptly varied around $N_\mathrm{d}\sim80$, 
we fit the plots with two functions according to the ranges of $N_\mathrm{d}$. 
In $N_\mathrm{d}\leq80$, the plots are fitted by the power law of $f_\mathrm{dif,min}=a_0\times N_\mathrm{d}^{b_0}$. 
On the other hand, 
we fit the plots by the function of $f_\mathrm{dif,min}=a_0$ in $N_\mathrm{d}\ge80$ 
because $f_\mathrm{dif,min}$ for $N_\mathrm{d}\ge80$ is almost constant. 
Although this constant $f_\mathrm{dif,min}$ is presumably caused by any factors, 
the cause of the existence of this lower limit for $f_\mathrm{dif,min}$ is not certain. 
Therefore, the cause of the lower limit is investigated in Section \ref{subsec_low}. 
In consequence, each plot in the right panel of Fig. \ref{df_nf_sph} is well fitted by 
\begin{eqnarray}
	f_\mathrm{dif,min}=
\left \{
\begin{array}{ll}
	6.5\times 10^2~N_\mathrm{d}^{-1.6}~\% ~~~~~\mathrm{for}~N_\mathrm{d}\leq80,\\
	0.8~~~~~~~~~~~~~~~~~~\%~~~~~\mathrm{for}~N_\mathrm{d}\ge80.
\end{array}
\right. 
\label{eq_bound}
\end{eqnarray}
Thus, the DF (template) for the isotropic model is recovered with about one percent error 
when the data number ($N_\mathrm{d}$) is larger than about several decades. 


\subsubsection{Anisotropic models}

\begin{figure}
\begin{center}
\includegraphics[width=85mm]{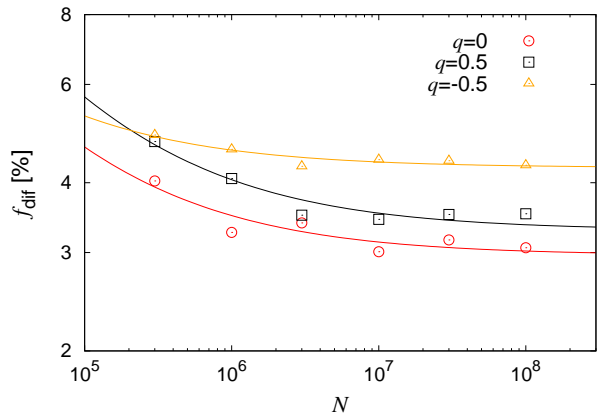}
\caption{
	Same as the left panel of Fig. \ref{df_nf_sph}, 
	but each plot is the case of the isotropic ($q=0$, red circle), radially anisotropic ($q=0.5$, black square), 
	and tangentially anisotropic ($q=-0.5$, orange triangle) target models. The data number is $N_\mathrm{d}=30$. 
}
\label{nf_sph_ani}
\end{center}
\end{figure}
We also show the dependence of ($f_\mathrm{dif}$) on the anisotropy of the target models. 
Fig. \ref{nf_sph_ani} represents $f_\mathrm{dif}$ as the function of $N$ for the different anisotropic models with $N_\mathrm{d}=30$. 
From Fig. \ref{nf_sph_ani}, 
$f_\mathrm{dif}$ for the anisotropic target models (green and blue plots) 
is about a few times larger than that for the isotropic target model. 
This is presumably due to the choice of the initial condition as mentioned in Section \ref{sec_ani}. 
The dependence of $f_\mathrm{dif}$ on the initial condition is investigated in Section \ref{subsec_ini}. 
From these results, 
the anisotropy for the spherical target models increases $f_\mathrm{dif}$ by a factor of about two 
when the initial condition is the Hernquist with Gaussian, which is the isotropic distribution.

\subsubsection{Three integrals models}


\begin{figure}
\begin{center}
\includegraphics[width=85mm]{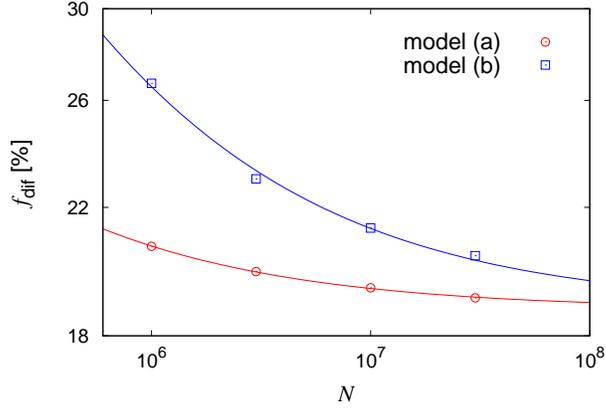}
\caption{
	The degree of accuracy for the recovery of the DFs $f_\mathrm{dif}$ for the axisymmetric three integral target models
	as a function of the particle number $N$. 
	Red circle and blue square plots represent the results for target models (a) and (b) of equations (\ref{ax_model}), respectively.  
	The mass data number ($N_\mathrm{m}$) is $32$, and the kinematic data number ($N_\mathrm{k}$) is $16$. 
	The red and green lines represent the curves fitted by $f_\mathrm{dif}=a_0\times N^{-0.5}+f_\mathrm{dif,min}$ for target models (a) and (b).
}
\label{nf_ani_ax}
\end{center}
\end{figure}
Fig. \ref{nf_ani_ax} shows $f_\mathrm{dif}$ as a function of $N$ 
for axisymmetric target models (a) and (b) of equations (\ref{ax_model}) 
with the mass data number ($N_\mathrm{m}$) of $32$ and kinematic data number ($N_\mathrm{k}$) of $16$. 
As a result, 
$f_\mathrm{dif,min}$ are $18.8\%$ for target model (a), and $18.9\%$ for target model (b). 
Thus, the different target models are reconstructed with the comparable degree of accuracy. 
From this result, 
we suppose that the target model (a) can be regarded as a representation of the axisymmetric three integral target model, 
and so we use the target model (a) below.

\begin{figure}
		\hspace{35mm}
\includegraphics[width=140mm]{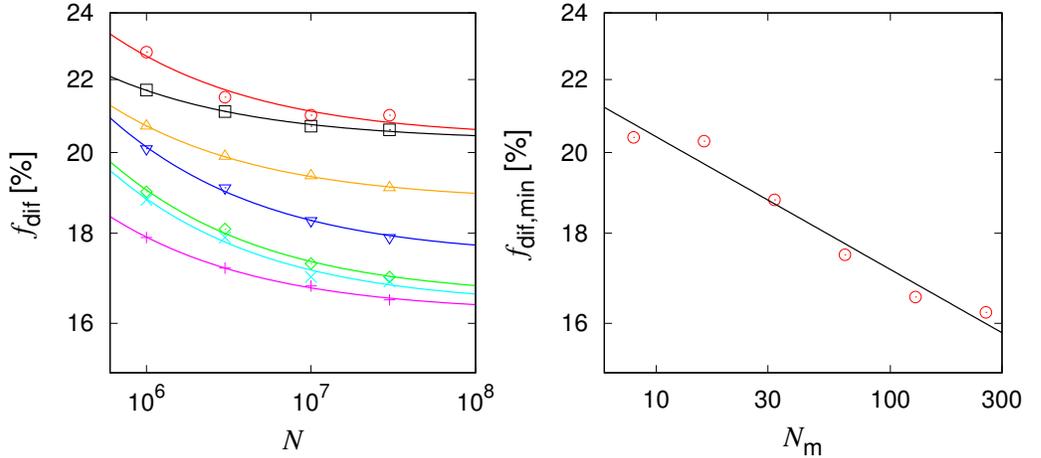}
\caption{
	The left panel shows the degree of accuracy for the recovery of the DFs $f_\mathrm{dif}$ 
	for the axisymmetric three integral target model of target model (a) as a function of the particle number $N$. 
	Red circle, black square, orange triangle, blue inverted triangle, green diamond, and magenta plus plots represent 
	the results for the kinematic data number $N_\mathrm{k}=16$ and
	the mass data number as $N_\mathrm{m}=$8, 16, 32, 64, 128, and 256, respectively. 
	Cyan cross plot represents the results for $N_\mathrm{k}=64$ and $N_\mathrm{m}=128$. 
	Each line represents the curve fitted by $f_\mathrm{dif}=a_0\times N^{-0.5}+f_\mathrm{dif,min}$ for concolorous plots. 
	The right panel shows $f_\mathrm{dif,min}$ as a function of $N_\mathrm{m}$ for $N_\mathrm{k}=16$. 
	The line represents the curve fitted by $f_\mathrm{dif,min}=a_0\times N_\mathrm{m}^{b_0}$.
}
\label{df_nf_ax}
\end{figure}

Next, we investigate the dependence of $f_\mathrm{dif}$ on $N$ and the data number 
for axisymmetric three integrals model (a). 
The left panel of Fig. \ref{df_nf_ax} shows $f_\mathrm{dif}$ as a function of $N$ for the several data number. 
As the cases of the spherical target model, 
we give the fitting function by $f_\mathrm{dif}=a_0\times N^{-0.5}+f_\mathrm{dif,min}$.
As can be seen from the fitted lines in Fig. \ref{df_nf_ax}, 
the convergences of $f_\mathrm{dif}$ to the minimum values ($f_\mathrm{dif,min}$) are in $N \sim 10^7 - 10^8$. 
On the other hand, the previous studies for the M2M method typically use the particle number from 
$N=5\times10^5$ to $1.8\times10^6$. 
In addition, the maximum particle number used in the M2M method is $N=6\times10^6$ \citep{del13}.
As also mentioned in the recovery of the DF for the spherical target model, 
the particle number that is required to recover the DFs with high accuracies 
is presumably larger than the particle number used by the previous studies for the M2M method. 
Meanwhile, the dependence of $f_\mathrm{dif}$ on $N_\mathrm{k}$ is weaker than that on $N_\mathrm{m}$ from the left panel of Fig. \ref{df_nf_ax}. 
This result suggests that the number of the kinetic grids 
are almost sufficient with $16\times 16$ to recover the DFs with this degree of accuracy. 

The right panel of Fig. \ref{df_nf_ax} shows $f_\mathrm{dif,min}$ as a function of $N_\mathrm{m}$ for $N_\mathrm{k}=16$. 
From this panel, 
the plots are well fitted by 
\begin{equation}
f_\mathrm{dif,min}=24.3~N_\mathrm{m}^{-0.075}~\%. 
\label{f_fit_ax}
\end{equation}
Thus, we derive the dependence of $f_\mathrm{dif,min}$ on $N_\mathrm{m}$ for the target model (a). 
If the relation of equation (\ref{f_fit_ax}) is valid at large $N_\mathrm{m}$, 
$f_\mathrm{dif,min}=15\%$ requires $N_\mathrm{m}\simeq600$, 
and $f_\mathrm{dif,min}=10\%$ requires $N_\mathrm{m}\simeq10^5$. 
However, 
because $f_\mathrm{dif,min}$ may have some lower limit in a way similar to the results for the spherical target model, 
one should be careful to use this relation at large $N_\mathrm{m}$.

\subsection{Initial condition dependence}\label{subsec_ini}

In the M2M modelling, the selection of the initial conditions 
significantly affects $f_\mathrm{dif}$. 
However, because the dependence of $f_\mathrm{dif}$ on the initial condition is not elucidated, 
we investigate the dependence in this section. 

\subsubsection{Isotropic models}

\begin{figure}
\begin{center}
\includegraphics[width=100mm]{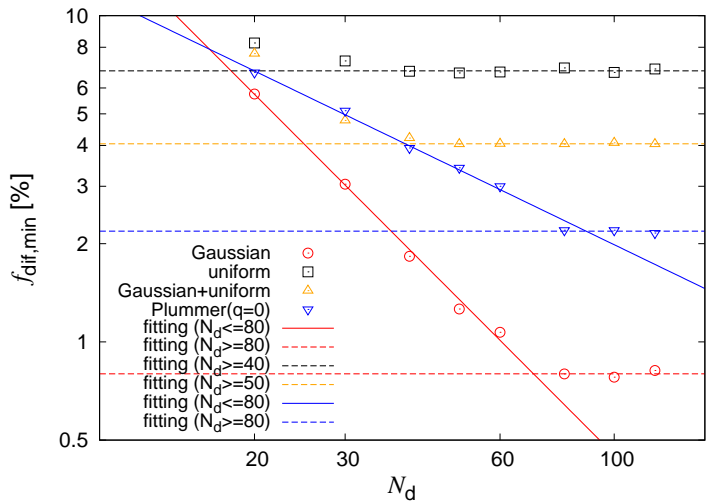}
\caption{Same as the right panel of Fig. \ref{df_nf_sph}, but the results for the several initial conditions. 
	Red circle, black square, orange triangle, and blue inverted triangle plots 
	represent the results for the initial conditions of the Hernquist with Gaussian, 
	the Hernquist with uniform, the Gaussian and uniform, and the isotropic Plummer model, respectively. 
	Red and blue lines are fitted curve with $f_\mathrm{dif,min}=a_0\times N_\mathrm{d}^{b_0}$ in $N_\mathrm{d}\leq80$ 
	for the Hernquist with Gaussian and the isotropic Plummer model. 
	Red, blue, orange, and black dashed lines are fitted with $f_\mathrm{dif,min}=a_0$ 
	in $N_\mathrm{d}\ge80$ for the Hernquist with Gaussian, 
	$N_\mathrm{d}\ge80$ for the isotropic Plummer model, 
	$N_\mathrm{d}\ge50$ for Gaussian and uniform, 
	and $N_\mathrm{d}\ge40$ for Hernquist with uniform, respectively. 
}
\label{df_sph3}
\end{center}
\end{figure}
We represent the dependence of $f_\mathrm{dif}$ on the initial condition for the isotropic Plummer target model. 
We use the following four initial conditions in Section \ref{subsec_ini}:
the Hernquist with Gaussian (also used in Section \ref{sec_5}), 
the Hernquist mass model with an uniform velocity distribution (Hernquist with uniform), 
the half particles are the Hernquist with Gaussian and the others are the Hernquist with uniform (Gaussian and uniform), 
and the isotropic Plummer model (same as the target model of Section \ref{sec_res_plu_iso} and this section). 
Fig. \ref{df_sph3} represents 
$f_\mathrm{dif,min}$ as a function of $N_\mathrm{d}$ for the four initial conditions. 
This figure indicates that $f_\mathrm{dif,min}$ for the Hernquist with Gaussian is lowest 
in the four initial conditions. 
The low value of $f_\mathrm{dif,min}$ for the Hernquist with Gaussian is presumably due to the relationship between 
the distribution of the particles and errors in the integrals of motion space. 
From the right panel of Fig. \ref{el_sph}, 
the normalized errors for the reconstruction of the isotropic spherical target model 
are high in the high- and low-energy regions. 
The particle distribution obeying the Hernquist with Gaussian 
has high density also in the high- and low-energy regions. 
Furthermore, the high error regions for the reconstruction 
will require more particles to recover the DFs more accurately. 
Therefore, we suppose that $f_\mathrm{dif,min}$ for the Hernquist with Gaussian is lowest of the four 
due to this concordance of the distributions in the integrals of motion space. 

On the other hand, from Fig. \ref{df_sph3}, $f_\mathrm{dif,min}$ for the Gaussian and uniform 
is lower than that for the Hernquist with Gaussian 
in spite of the distribution of the half particles for the Gaussian and uniform being the same as that for the Hernquist with Gaussian. 
This result implies that distributing the particles widely and finely in the three integrals of motion space 
is not sufficient for the M2M method to reconstruct the DFs accurately. 
This is because to addition the particles 
makes the recovery of the DFs less accurate from the results of the recovery for the Gaussian and uniform. 
Furthermore, we find from Fig. \ref{df_sph3} that $f_\mathrm{dif,min}$ are limited 
by respective values according to the initial conditions. 
The various values of the lower limits imply that 
the lower limit is not caused by a numerical error 
because a numerical error gives a certain value for a lower limit irrespective of the initial conditions. 
However, since the cause of the appearance of the lower limits is not certain, 
we try to identify the cause of the appearance of the lower limits in Section \ref{subsec_low}. 

\begin{figure}
	\begin{center}
\includegraphics[width=120mm]{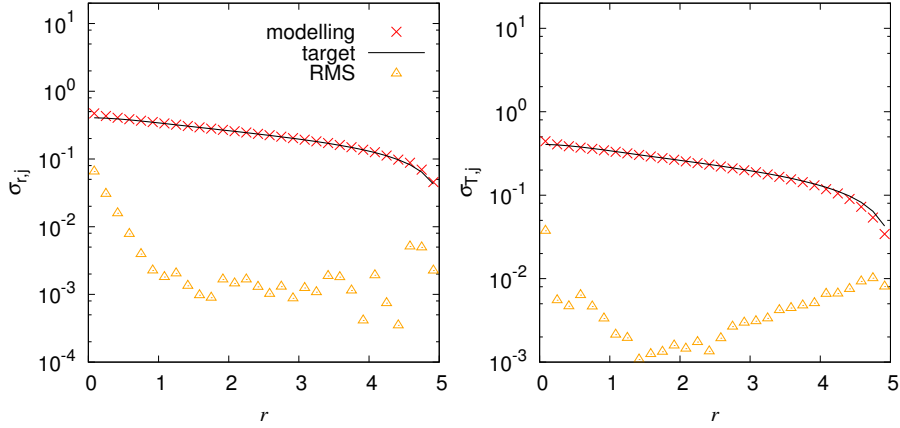}
\caption{Same as Fig. \ref{vel_sph}, but the initial condition is the Hernquist with uniform.}
\label{vel_sph_30}
\end{center}
\end{figure}
To find the characteristics of the better initial condition for the high accurate recovery of the DFs, 
we investigate the recovery of the velocity distribution for the Hernquist with uniform. 
Fig. \ref{vel_sph_30} shows the recovery of the velocity dispersion distribution of the isotropic target model 
for the Hernquist with uniform with $N_\mathrm{d}=100$. 
As a result, $\overline{\mathrm{RMS}}(\sigma_r)$ and $\overline{\mathrm{RMS}}(\sigma_\mathrm{T})$ are $2.13$ and $3.49\%$, respectively. 
These values of the Hernquist with uniform are higher than those of the Hernquist with Gaussian as the recovery of the DFs. 
As can be seen from Fig. \ref{vel_sph_30} and \ref{vel_sph}, 
the RMS values of the inner region ($r<1$) for the Hernquist with uniform are higher than that for the Hernquist with Gaussian. 
From this result, 
the reason of the lower limit of the recovery of the DF is probably 
due to the worse recovery of the velocity distribution in this region. 
However, since it is difficult to find the better initial condition for the accurate recovery, 
we set finding the better initial condition as a future work. 
	On the other hand, the problem of a poorly chosen initial condition may be mitigated by a resampling scheme 
	such as is implemented in \citet{deh09} and \citet{hun14a}, 
	which increases the number of particles in the phase-space regions of high weights. 
	Nevertheless, to find the best resampling scheme for this purpose is presumably difficult
	because it is related to the problem that what kind of the initial condition is better for the accurate reconstruction of the DFs. 
	Therefore, we also set finding the better way of the resampling scheme as a future work. 


\subsubsection{Three integrals models}

\begin{figure}
\begin{center}
\includegraphics[width=90mm]{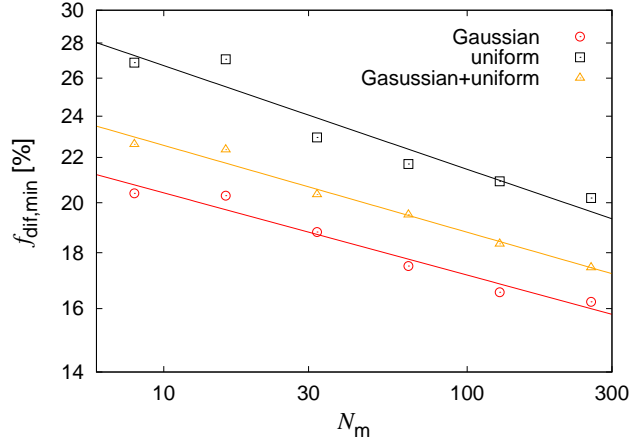}
\caption{
	Same as the right panel of Fig. \ref{df_nf_ax}, but for the several initial conditions. 
	The red circle, black square, and orange triangle plots represent the results for 
	the initial conditions of the Hernquist with Gaussian, the Hernquist with uniform, and the Gaussian and uniform, respectively.
	}
\label{df_sph4}
\end{center}
\end{figure}
Here we investigate the dependence of $f_\mathrm{dif,min}$ 
on the initial condition for axisymmetric target model (a) with $N_k =16$. 
Fig. \ref{df_sph4} shows $f_\mathrm{dif,min}$ as a function of $N_\mathrm{m}$ 
for the three initial conditions (Hernquist with Gaussian, Hernquist with uniform, and Gaussian and uniform). 
As a result, 
$f_\mathrm{dif,min}$ for the Hernquist with Gaussian is lowest 
in the three initial conditions. 
Here the large errors for the recovery of the DF for the axisymmetric target model (a) 
mainly appear in the high- and low-energy regions as seen from Fig. \ref{4_el}. 
Therefore, as described in the spherical cases, 
the relationship of the distribution between the high errors and the high particle density 
is probably key to choose the initial condition that constructs the DFs accurately. 
Although the lower limits of $f_\mathrm{dif,min}$ are not observed against the spherical cases, 
it is not clear whether the lower limits of $f_\mathrm{dif,min}$ exist or not 
for the recovery of the axisymmetric target model. 
To investigate the characteristics of the lower limits, 
we search for the cause of the existence of the lower limits in the next section. 

\subsection{Search for the cause of the lower limit}\label{subsec_low}

We see from the right-hand panel of Fig. \ref{df_nf_sph} and Fig. \ref{df_sph3} that 
the degree of accuracy for the recovery of the DFs for the isotropic spherical target models 
is limited by causes. 
From the results in these figures, 
the cause is not a numerical error, a shortage of the particle number and the data number. 
Since the cause of the existence of the lower limits is not obvious, 
we try to identify the cause in this section. 

We set that a target model is the isotropic Plummer model ($q=0$), 
the initial condition is the Gaussian and uniform, $N_\mathrm{d}=100$, 
and $N=10^5$, $3\times 10^5$, $10^6$, $3\times10^6$, $10^7$, and $3\times10^7$ 
because these parameters are the conditions whose results, which are shown in Fig. \ref{df_sph3}, suffer the effect of the lower limits. 
We search for the cause of the lower limit in the following way: 
By constructing the target model with the several conditions described below (additional conditions), 
we derive $f_\mathrm{dif,min}$ in the same manner as shown in section \ref{sec_dis_dep_iso}. 
We compare derived $f_\mathrm{dif,min}$ with the additional conditions to 
$f_\mathrm{dif,min}$ ($=4.2~\%$) without the additional conditions. 
If both $f_\mathrm{dif,min}$ accord to each other, 
we regard the additional conditions as not cause the lower limit. 
We investigate the additional conditions as 
higher order velocity moments, 
the entropy parameter, 
temporal smoothing effect, and the configuration of the kinematic observables. 

\subsubsection{Higher-order velocity moments}


We first investigate whether the absence of the higher-order velocity moments causes the lower limits. 
We derive $f_\mathrm{dif,min}$ using the kinematic observable until the $6$th order velocity moments. 
The velocity contribution parameters $\lambda_{h_5}$ and $\lambda_{h_6}$ are set to be $0.05$. 
As a result, $f_\mathrm{dif,min}$ for the reconstruction with the higher-order velocity moments (5th and 6th order) is $4.2$ $\%$. 
Since $f_\mathrm{dif,min}$ without the higher-order velocity moments is also $4.2$ $\%$, 
the absence of the higher-order velocity moments is not the cause of the lower limit.

\begin{figure}
	\begin{center}
	\includegraphics[width=90mm]{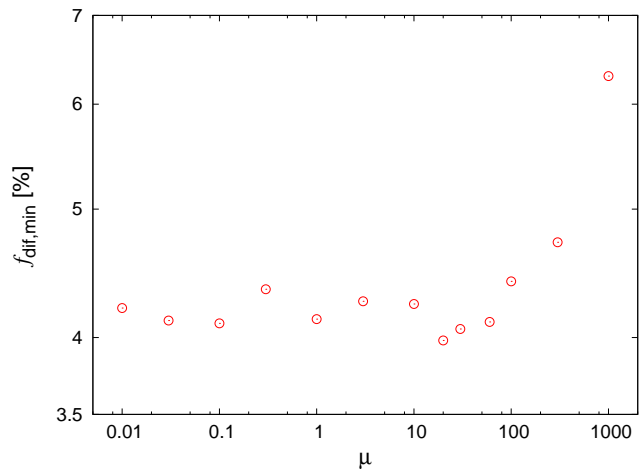}
	\caption{
		Same as the right panel of Fig.\ref{df_nf_ax}, but as a function of $\mu$ for $N_\mathrm{d}=100$. 
	}
	\label{mu_sph}
	\end{center}
\end{figure}
\begin{figure}
	\begin{center}
	\includegraphics[width=160mm]{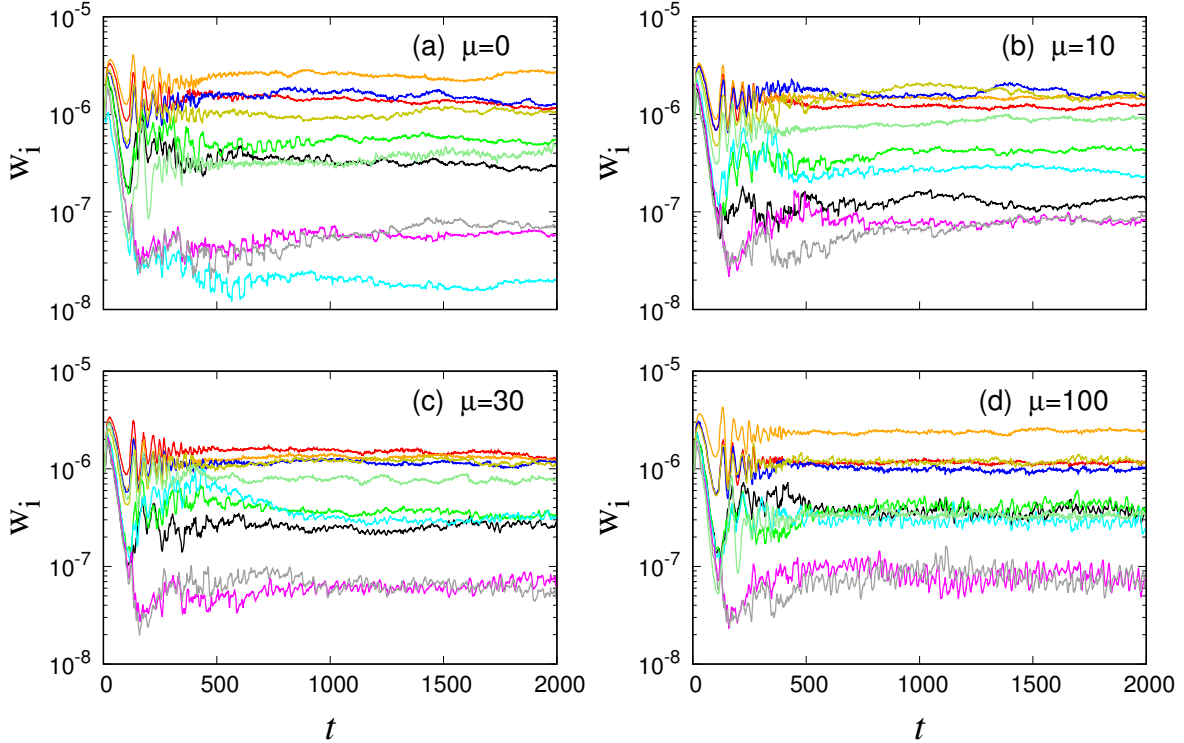}
	\caption{
		The weight evolution for randomly selected ten particles. 
		The entropy parameter is $\mu=$0 (a), 10 (b), 30 (c), and 100 (d), respectively. 
		The particle number is $N=10^6$, and the data number is $N_\mathrm{d}=100$. 
	}
	\label{weight_ev}
	\end{center}
\end{figure}
\subsubsection{Regularization}
In this paper, the entropy parameter $\mu$ is set to be 0. 
However, without the regularization term, the particle weights do not actually converge. 
Previous studies \citep*{sye96,del07,lon10,hun13} all find the choice of $\mu$ to be important for convergence of the model. 
Furthermore, the several studies \citep*{del08,mor12,mor13,hun13} indicate that 
the regularization term with appropriate values of $\mu$ makes the recovery of the observables better. 
To investigate the effects of the regularization term on the degree of accuracy for the recovery of the DFs, 
we change the entropy parameter from $10^{-2}$ to $10^3$. 
Fig. \ref{mu_sph} shows $f_\mathrm{dif,min}$ as a function of $\mu$. 
From this figure, $f_\mathrm{dif,min}$ is slightly decreased around $\mu \sim 30$. 
This decreasement is consistent with the previous studies that 
appropriate values of $\mu$ is a bit smaller than the values that the recovery becomes worse. 
Fig. \ref{mu_sph} also indicates that the effect of the regularization on $f_\mathrm{dif,min}$ is not significant in this settings. 

Next, to verify the degree of convergence due to the regularization term, we investigate the behavior of the weights according to $\mu$. 
Four panels of Fig. \ref{weight_ev} show the weight evolution for $N=10^6$, and $N_\mathrm{d}=100$ according to several $\mu$. 
As seen in (d) of Fig. \ref{weight_ev}, 
the weights that have low values strongly oscillate until $t=2000$ when the entropy has high value ($\mu=100$). 
Thus, the convergence is not well accomplished at large $\mu$ due to the overregularization. 
In the lower entropy cases of $\mu=0$ (a) and $\mu=10$ (b), 
the weights that have low values (e.g. grey line) fluctuate in long periods. 
Hence, the entire weights are not converged well in the lower entropy cases. 
In the intermediate entropy case of $\mu=30$ (c), 
the entire weights are converged well compared with the other cases, 
although the complete convergence of weights is thought to be not yet accomplished. 
On the other hand, \citet*{mor12} introduced the new regularization method, 
which gives the prior in equation (\ref{regularization_term}) by the averages among neighbor weights in integrals of motion space. 
This new method possibly makes the recovery of the DFs better. 
Furthermore, the improvement 
for the temporal smoothing \citep{mal15} 
investigated in next section also makes convergence of weights much better. 

\subsubsection{Temporal smoothing effect}\label{subsubsec_smo}

Since the temporal smoothing is finite in the M2M method, 
this finiteness may cause the lower limits. 
Therefore, 
we investigate the effect of the shortage of the temporal smoothing on the lower limit. 
Recently, \citet{mal15} 
introduce a new development for the M2M method. 
They give the kernel as time average occupancy of each particle in each observable grid, 
and so this method removes the shortage of the temporal smoothing similar to the orbit-based method. 
We use this procedure to remove the finiteness of the temporal smoothing. 
We calculate the time average occupancy for $10^5$ steps (1000 units). 
The result of the reconstruction shows that $f_\mathrm{dif,min}$ with the procedure is $4.1~\%$. 
This result indicates that the lower limit does not result from the shortage of the temporal smoothing.  

We also investigate whether the deficiency of the resolution causes the lower limits. 
The inner boundary radius of mass $r_\mathrm{min}=10^{-4}$ and kinematics $r_\mathrm{pmin}=10^{-4}$ 
may be not sufficiently small to recover the DFs accurately.
Therefore, we use the $r_\mathrm{min}=10^{-6}$ and $r_\mathrm{pmin}=10^{-6}$. 
Since the temporal smoothing is especially important for the accurate calculation of such small regions, 
we additionally use the procedure introduced by \citet{mal15}. 
The result shows that $f_\mathrm{dif,min}$ is 4.1 $\%$ 
for the reconstruction with $r_\mathrm{min}=10^{-6}$ and $r_\mathrm{pmin}=10^{-6}$. 
Hence, the deficiency of the resolution of the inner region does not lead to the lower limit.

\subsubsection{Configurations of the kinematic observables}\label{subsubsec_deg}

Up to here, we use the LOSVD 
as kinematic observables. 
Since the shortage of kinematic information possibly causes the lower limits, 
we change the configuration of the kinematic grids of the observational data. 
We assume the case that the LOSVD can be observed from multiple directions 
to reduce the shortage of the kinematic information. 
We give the kinematic information seen from three directions. 
One is the same as the normal LOS direction, 
and the other two directions are perpendicular to the normal LOS direction and perpendicular to each other. 
The result of the reconstruction with the three directional kinematic observables shows $f_\mathrm{dif,min}=4.2~\%$. 
Thus, the insufficiency of the directions of the kinematic information does not cause the lower limit. 

To investigate the effect of the projection of the LOSVD on the $f_\mathrm{dif,min}$, 
we cut the kinematic grids perpendicular to the LOS direction at equal intervals. 
We set 
the number of the kinematic grids in the LOS direction ($N_\mathrm{LOS}$) as 2, 10, 30, and 100. 
From the results of the reconstructions with $N_\mathrm{LOS}=$1, 2, 10, 30, and 100, 
$f_\mathrm{dif,min}$ are 4.2, 3.4, 3.4, 3.3, and 3.3$\%$, respectively. 
Although $f_\mathrm{dif,min}$ is a little reduced by the increment of $N_\mathrm{LOS}$, 
$f_\mathrm{dif,min}$ is again limited around $N_\mathrm{LOS}\sim 2$. 
Consequently, the lower limit is almost unchanged by removing the degeneration along the LOS direction.

Thus, the reason for the lower limit of the $f_\mathrm{dif,min}$ remains unresolved. 
We suppose that the problem for the lower limit is related to the way of the M2M method. 
In the M2M method, the weights are evolved by solving the equation (\ref{eq:foc}). 
However, because 
the way the weights are evolved is not unique, 
a better way to evolve the weights will be found. 
Therefore, the improvement for the M2M method may be required to solve the problem. 

\subsection{Future observations}

Recently, 
an era promising great progress in astrometry has begun. 
Gaia was launched on $2013$ December $19$ and began routine operations in $2014$ August. 
Gaia has the aim of mapping more than a billion stars ($V~\leq~20$) in our Galaxy. 
Gaia measures parallaxes with an accuracy of $5-25~\mu$as, positions with an accuracy of $4-19~\mu$as 
and proper motions with an accuracy of $3-13$ $\mu$as/year 
for stars brighter than $V=15$ mag \citep{per01}. 
	Gaia will also provide the spectroscopic radial velocity measurements for about 150 million stars. 
	The expected data release dates for Gaia are 14 September 2016, 2017, 2018, 2019, and 2022. 
	The data for the parallaxes, the proper motions, and the radial velocities 
	are released from the second release in 2017. 
Small-JASMINE measures parallaxes, positions with an accuracy of $\sim 10$ $\mu$as 
and proper motions with an accuracy of $\sim 10$ $\mu$as/year 
for stars brighter than Hw (1.1$\sim1.7~\mathrm{\mu m}$)$~=~12$ mag. 
Small-JASMINE will observe stars towards the Galactic nucleus bulge 
around the center of the bulge of our Galaxy \citep{jas}. 
It is supposed that Small-JASMINE will be launched around $2022$.
Combining the astrometry with the spectroscopic observations, which provide radial velocities, 
we will directly obtain the six-dimension phase space coordinates of observed stars. 

Using these observational data, 
we will determine the dynamical structure accurately. 
Here the dominant component of the errors 
for the decision of the phase space coordinates or the values of integrals of motion of observed stars is parallaxes. 
For the stars whose distances are ten $\mathrm{kpc}$, 
the error of the distance is about ten percent, 
the error of the position for the directions of the right ascension and the declination is about 0.1 AU, 
the error of the proper motion is about 0.2-0.5 km/s, 
and the error of the radial velocity in the Gaia spectroscopic measurements is 
about 1-15 km/s. 
Therefore, the uncertainties of the observed six-dimension phase space coordinates of stars 
are supposed to be also about ten percent. 
However, the uncertainties of the templates (DFs) 
for the axisymmetric three integrals target model 
are about a few tens percent according to our results. 
Hence, we suggest that the degree of accuracy for the recovery of the dynamical structure 
may be limited by the uncertainties of the templates. 

	Meanwhile, in recent study of \citet{por15b}, they constructed a variety of templates for the Milky Way bulge/bar using the M2M method 
	and derived the fraction of orbit classes. 
	However, our results imply that derived DFs have large uncertainties 
	and the fraction of orbit classes may also have large uncertainties. 
	Also, the M2M method in \citet{deg10} and \citet{hun13,hun14b} 
	calculates the gravitational potential via self-gravity of the model particles. 
	Such modelling can reduce the parameters of the galactic model and the number of the templates that should be prepared. 
	Furthermore, this modelling may improve the recovery of the DFs because of the self-consistency. 
	On the other hand, this modelling also has disadvantages 
	such as the difficulty of weight convergence, and substantial computational time. 
	Since it is important to investigate the performance of the M2M method using the self-consistent model, 
	we set this as a future work. 